\documentclass[11pt,a4paper]{article}

\usepackage[T1]{fontenc}
\usepackage[utf8]{inputenc}
\usepackage{lmodern}        % high-quality font
\usepackage{microtype}      % improves typography
\usepackage{geometry}
\newcommand{\EQ}{\begin{equation}}
\newcommand{\EN}{\end{equation}}
 
\usepackage[english]{babel}

\usepackage{amsmath}
\usepackage{amssymb}
\usepackage{amsfonts}
\usepackage{mathtools}
\usepackage{mathrsfs}
\usepackage{bm}
\usepackage{braket}
\usepackage{esint}
\usepackage{ytableau}

\usepackage{graphicx}
\usepackage{subcaption}
\usepackage{caption}
\usepackage{booktabs}
\usepackage{array}
\usepackage{multirow}

\usepackage{xcolor}

\usepackage[
    colorlinks=true,
    linkcolor=blue,
    citecolor=blue,
    urlcolor=blue
]{hyperref}

\usepackage[
    style=phys,
    sorting=none,
    eprint = true
]{biblatex}

\DeclareFieldFormat[article]{title}{\mkbibemph{#1}}
\usepackage{tikz}
\usetikzlibrary{quantikz2}

\usepackage{titlesec}

\titleformat{\section}
{\large\bfseries}
{\thesection.}{0.5em}{}

\titleformat{\subsection}
{\normalsize\bfseries}
{\thesubsection.}{0.5em}{}

\usepackage{etoolbox}

\newif\ifappendix
\appendixfalse

\pretocmd{\appendix}{%
    \appendixtrue
    \renewcommand{\thesection}{\Alph{section}}
}{}{}

\newcommand{\dd}{\mathrm{d}}

\title{\bfseries Quantum Stochastic Walks on the Permutation Group}

\author{
Feng He,
Arthur Hutsalyuk,
Giuseppe Mussardo,
Andrea Stampiggi\thanks{Contact author: astampig@sissa.it}
\\[5pt]
\small
International School for Advanced Studies (SISSA),
Via Bonomea 265, 34136 Trieste, Italy \\
\small
INFN Sezione di Trieste, via Valerio 2, 34127 Trieste, Italy
}

\tiny\date{\today}

\begin{document}

\maketitle

%%%%%%%%%%%%%%%%%%%%%%%%%%%%%%%%%%%%%%%%%%%%
%% ABSTRACT
%%%%%%%%%%%%%%%%%%%%%%%%%%%%%%%%%%%%%%%%%%%%

\begin{abstract}
\noindent
How rapidly does order give way to randomness, and can quantum coherence accelerate this process? We address these questions through the paradigmatic problem of card shuffling, formulated as a random walk on the symmetric group $S_n$.
We first recast the random-transposition walk studied by Diaconis and Shahshahani, as well as more general walks generated by conjugacy classes of $S_n$, in continuous time. We then identify the transition matrix of each classical walk with a permutation Hamiltonian generating a corresponding unitary quantum walk.
Purely unitary evolution, however, does not generically converge to the uniform distribution in the classical sense of mixing: coherence preserves information rather than erasing it. 
We therefore embed the problem into a quantum stochastic walk, where coherent dynamics competes with the dissipative process responsible for classical mixing. 
In this setting, quantum coherence assists randomization. We prove that it can only decrease the distance from the uniform distribution in the computational basis and can therefore accelerate mixing. An  analysis of the slowest mode yields a criterion for the coupling strength required to produce an appreciable speedup. Finally, numerical results reveal a scaling collapse of the ratio between quantum and classical mixing times onto a simple one-parameter form. Our results illustrate how coherence and dissipation can cooperate in the emergence of randomness in walks on permutation groups. 
\end{abstract}

%%%%%%%%%%%%%%%%%%%%%%%%%%%%%%%%%%%%%%%%%%%%
%% CONTENT
%%%%%%%%%%%%%%%%%%%%%%%%%%%%%%%%%%%%%%%%%%%%

\section{Introduction}

Card shuffling provides one of the simplest and most transparent realizations of a general problem: how does an ordered system lose memory of its initial condition and become random? At first sight, this question may seem to belong more naturally to the realm of recreational mathematics rather than to fundamental science. Yet few problems have proven so influential in probability theory \cite{Diaconis_1988, Levin2017}, statistical mechanics \cite{Martinelli2004}, computer science \cite{Sinclair1993} and group theory \cite{SaloffCoste2004}. 

This question can be formulated naturally in terms of random walks on the symmetric group $S_n$. Starting from the identity permutation, one repeatedly applies random elementary operations drawn from a prescribed probability distribution. The resulting stochastic process defines a Markov chain whose equilibrium distribution, under appropriate ergodicity conditions, is uniform over the accessible permutations. The central question is then how rapidly the walk approaches this equilibrium distribution.

A major breakthrough in the understanding of this problem was achieved by Diaconis and Shahshahani (DS)\cite{Diaconis_1981}, who showed that the approach to randomness exhibits a remarkable cutoff phenomenon. For the random-transposition shuffle, the distance from the uniform distribution  drops from near its maximal value to near zero within a narrow time window. Thus, rather than losing memory gradually, the system undergoes a sharp crossover from order to randomness. This result has become a paradigmatic example of cutoff in finite Markov chains and has profoundly influenced the modern theory of mixing and random walks on groups.

The notion of ``shuffling'', however, is far from unique. Within the permutation group, one can generate a plethora of elementary moves and probability distributions: all-to-all transpositions, adjacent exchanges, or more general class operations involving cycles of fixed length, to say a few. Provided that the resulting walk is ergodic on its accessible state space, it approaches the uniform distribution on that space at long times. Nevertheless, different choices of elementary moves can lead to dramatically different mixing behavior. The different probability distributions leave a distinct fingerprint on the corresponding randomization process; such distributions can be discriminated, for instance by inspecting the extremely regular structure for the group, such as that of the irreducible representations and characters.

The development of quantum technologies naturally raises a broader question: if the same permutations are implemented through coherent quantum dynamics rather than a classical stochastic process \cite{Kempe2003,Mlken2011,VenegasAndraca2012}, how efficiently can the resulting dynamics generate randomness? Also, how are classical mixing and quantum equilibration related when both are generated by the same elementary operations on the symmetric group?

More broadly, equilibration in isolated quantum systems is a central problem in nonequilibrium many-body physics \cite{Gogolin}. Following a quantum quench \cite{Calabrese_2007}, unitary dynamics may lead to the equilibration of observables through dephasing, even though the global state never converges to a stationary state. Depending on the integrability and spectral structure of the system, late-time observables may be described by conventional thermal ensembles, generalized Gibbs ensembles, or may fail to thermalize altogether \cite{CalabreseEsslerMussardo2016,RevModPhys.83.863,Huse-review,Moudgalya2022}. Random walks on $S_n$ provide a particularly transparent setting in which to examine the relation between classical mixing and quantum equilibration, because the classical transition matrix can also be interpreted as the Hamiltonian generating the corresponding quantum walk.

In this work, we formulate and compare classical and quantum walks on $S_n$ generated by conjugacy classes of $q$-cycles. For $q=2$, this construction reduces to the random-transposition walk studied by DS. The normalized class operator that acts as the transition matrix of the classical continuous-time random walk (CCTRW) can also be interpreted as the Hamiltonian generating the corresponding unitary quantum walk. Despite sharing the same underlying operator, the classical and quantum walks produce probability distributions with fundamentally different long-time behavior. 

In the classical setting, mixing refers to convergence toward the uniform distribution \cite{Diaconis_1981,Diaconis_1988}. By contrast, unitary evolution preserves information and does not generally cause the computational-basis measurement probabilities to become stationary. Starting from a computational-basis state, the quantum walk produces a uniform measurement distribution at a given time only if the evolved state is a superposition of computational basis states with equal amplitude modulus, although their relative phases may differ. Moreover, even the long-time-averaged distribution need not be uniform. Quantum mixing times are therefore commonly defined through the convergence of finite-time-averaged measurement distributions to their long-time limits \cite{chakraborty2020fast}, rather than through convergence of the instantaneous distribution to uniformity. Consequently, classical and unitary mixing describe intrinsically different notions of equilibration and cannot in general be compared directly.

This distinction reflects two fundamentally different mechanisms for exploring $S_n$. Classical stochastic dynamics progressively erases information about the initial configuration as the probability distribution approaches equilibrium. Unitary dynamics, by contrast, preserves the information contained in the full quantum state while spreading its amplitudes across the Hilbert space through coherent interference. To quantify this difference, we study diagnostics such as the inverse participation ratio and the spectral variance. These quantities define a quantum dephasing timescale, characterizing the onset of coherent spreading, which we compare with the mixing time of the CCTRW generated by the same class operator. To recover genuine mixing in a quantum setting, one must introduce irreversible dynamics that erase information. 

In the second part of this work we consider quantum stochastic walks (QSWs) on $S_n$ \cite{PhysRevA.81.022323}, described by Lindblad dynamics combining coherent evolution with Markovian dissipation. A similar setup has been considered for the case of open quadratic fermionic dynamics, with an emphasis on mixing times and the cut-off phenomenon  \cite{SciPostPhys.9.4.049}. We initialize the density matrix $\rho$ in a pure computational-basis state and construct the dissipative part of the evolution so that, in the absence of coherent dynamics, its diagonal entries obey the CCTRW master equation. Under the appropriate ergodicity conditions, $\rho$ then converges to the maximally mixed state (MMS), and its computational-basis probabilities converge to the uniform distribution.

For dissipators generated by uniformly sampled $q$-cycles and coherent evolution generated by an arbitrary class Hamiltonian, we show that the trace distance between the evolving density matrix and the maximally mixed state is independent of the coherent coupling. By contrast, the distance between the computational-basis probability distribution and the uniform distribution does depend on the coherent dynamics. We prove that this distance is never greater than in the corresponding classical walk. Thus, although coherence does not accelerate convergence of the full density matrix in trace distance, it can accelerate mixing of the measurement probabilities. 
Our central result can be stated as follows. For a dissipator generated by uniformly sampled \(q\)-cycles and any coherent class Hamiltonian \(H_m\), the trace distance between the evolving density matrix and the maximally mixed state is independent of the coherent coupling \(\alpha\). By contrast, the total-variation distance of the computational-basis probabilities from the uniform distribution satisfies
$$ D_\alpha(t)\le D_0(t) $$
for all \(t\). Thus coherence cannot accelerate the loss of information at the level of the full density matrix, but it can redistribute part of that information into off-diagonal coherences, thereby making the measured probability distribution more uniform than in the corresponding classical walk. It is important to stress that this result does not hold for arbitrary coherent dynamics. Its validity relies on the special algebraic structure considered here: the dissipative dynamics is generated by a conjugacy class of \(q\)-cycles, while the coherent dynamics is generated by a class Hamiltonian \(H_m\). Since class operators belong to the center of the group algebra, the coherent and dissipative superoperators commute, leading to the factorization of the dynamics underlying the above inequality. For a generic noncentral permutation Hamiltonian this commutativity is lost, and no analogous monotonic enhancement of mixing is guaranteed.
An analysis of the slowest-decaying mode yields a criterion for the coherent coupling required to produce an appreciable speedup. Finally, numerical results for $n\leq 23$ reveal a finite-size scaling collapse of the ratio between quantum and classical mixing times, consistent with a simple one-parameter form.

The paper is organized as follows. Section~\ref{s_DS} reviews the DS theorem for random transpositions and its generalization to random $q$-cycles. In Section~\ref{s_classical}, we formulate the corresponding classical randomization process as a CCTRW and discuss both its instantaneous and time-averaged probability distributions. Section~\ref{s_quantum} introduces the associated unitary quantum walks generated by permutation Hamiltonians and discusses their dephasing properties. In Section~\ref{s_quantum_stochastic}, we connect the classical and unitary settings within the framework of quantum stochastic walks by combining a $q$-cycle dissipator with coherent evolution generated by an $m$-cycle Hamiltonian. Section~\ref{s_conclusions} summarizes our conclusions and discusses the interplay among coherence, dissipation, and the emergence of randomness. Additional technical material is presented in the appendices: Appendix~\ref{a_lower_bound_coupon_collector} derives a lower bound on the mixing time for random $q$-cycles, while Appendix~\ref{a_IPR_class} analyzes the IPR of class operators, with particular emphasis on dephasing and quantum recurrence timescales.

\section{The Diaconis--Shahshahani theorem}
\label{s_DS}
A natural benchmark for any discussion of quantum shuffling is the classical random-transposition shuffle. Consider an initially ordered collection of $n$ distinct objects, such as balls of different colors. At each step, two labels $L,R\in\{1,\ldots,n\}$ are chosen independently and uniformly. If $L\neq R$, the corresponding objects are transposed, whereas if $L=R$, the configuration remains unchanged. The central question is how many steps are required for the resulting probability distribution over permutations to become close to uniform within a set tolerance $\epsilon$.

DS showed \cite{Diaconis_1981} that this walk exhibits a sharp cutoff around a number of shufflings equal to 
\begin{equation}
\label{nlogn}
k^* \sim \frac{1}{2}n\log n.
\end{equation}
More precisely, the distance from the uniform distribution drops sharply within a window $O(n)$ around the leading timescale $n\log n$. Before stating this result more precisely, we introduce a few basic notions concerning probability measures on finite groups. It is already worth emphasizing the striking separation of scales: the number of possible configurations grows as $n!$, whereas only $O(n\log n)$ steps are required for the distribution to become close to uniform.

\begin{figure}[t]
\centering
\includegraphics[width=0.75\textwidth]{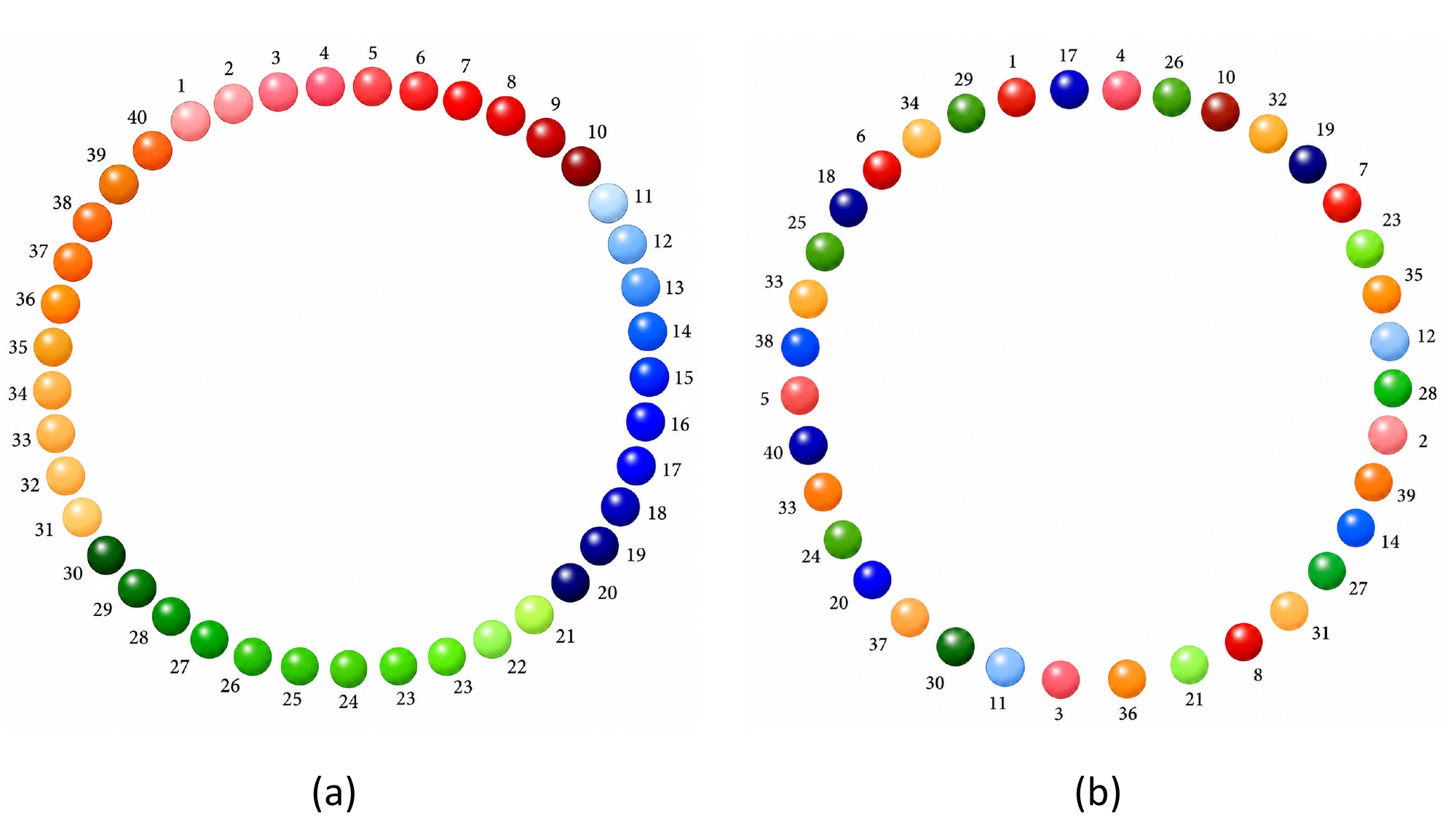}
\caption{{(a) Ordered and (b) Random configurations of 40 balls}}
%\ref{tablech} .) }
\label{propvert}
\end{figure}

\paragraph{Probability Measures and Total Variation Distance.}
Let $G$ be a finite group of order $|G|$. A probability measure on $G$ is a function $P:G\to[0,1]$ satisfying
\begin{equation}
    \sum_{g\in G}P(g)=1.
\end{equation}
Thus, $P(g)$ is the probability assigned to the group element $g$. Given two probability measures $P$ and $Q$ on $G$, their total variation (TV) distance is
\begin{equation}
\label{total-variance}
    D(P,Q)
    =\|P-Q\|_{\mathrm{TV}}
    =\frac{1}{2}\sum_{g\in G}|P(g)-Q(g)|.
\end{equation}
This distance quantifies the distinguishability of the two distributions. It satisfies $0\leq D(P,Q)\leq 1$: it vanishes if and only if $P=Q$, while it equals one if the two distributions have disjoint supports.

Other distances can also be used to study convergence to equilibrium. For example, define the squared $L_2$ distance by
\begin{equation}
\label{eq_L_2_norm}
    D_2(P,Q)
    =\|P-Q\|_2^2
    =\sum_{g\in G}|P(g)-Q(g)|^2.
\end{equation}
The TV and squared $L_2$ distances satisfy
\begin{equation}
\label{bounds_uniform}
    D_2(P,Q)
    \leq 4D(P,Q)^2
    \leq |G|D_2(P,Q).
\end{equation}
The first inequality follows from $\|x\|_2\leq\|x\|_1$, whereas the second follows from the Cauchy--Schwarz inequality.

For the problem considered by DS, $G=S_n$, the symmetric group of $n$ objects, whose order is $|S_n|=n!$. The target distribution is the uniform measure $U(g)=1/n!$. The variables $L$ and $R$ label the objects to be exchanged, and $P_{L,R}$ denotes their transposition. The probability of selecting $L=R$, in which case the identity permutation $e$ is applied, is therefore $T(e)=1/n$. When $L\neq R$, each transposition $\tau$ occurs with probability $T(\tau)=2/n^2$. Indeed, there are $N_2=n(n-1)/2$ distinct transpositions, all equally likely, and the total probability of applying a nonidentity move is $1-T(e)=(n-1)/n$. For every permutation $\sigma$ that is neither the identity nor a transposition, $T(\sigma)=0$. Thus, $T$ is a nonnegative, normalized probability measure.

The essential property of the DS measure is that all transpositions are equally likely. Consequently, the measure is constant on conjugacy classes. We will later generalize this construction to probability measures supported on permutations of $q$ objects.

\paragraph{Random Walks on Groups and Convolution.}
The probability measure $T$ induces a transition matrix on the configuration space $S_n$:
\begin{equation}
    M_{g\to h}=T(hg^{-1}).
\end{equation}
The matrix $M$ is bistochastic, meaning that
\begin{equation}
    \sum_g M_{g\to h}
    =
    \sum_h M_{g\to h}
    =1,
\end{equation}
and it defines a random walk on $S_n$. From a group-theoretic perspective, $M$ acts in the regular representation of $S_n$. Starting from $g$, the probability of reaching $h$ after $k$ steps is $[M^k]_{g\to h}$, which is given by the $k$-fold convolution of $T$. For two probability measures $P_1$ and $P_2$, their convolution is defined by
\begin{equation}
    (P_2*P_1)(\gamma)
    =
    \sum_{\sigma\in S_n}
    P_2(\gamma\sigma^{-1})P_1(\sigma).
\end{equation}
Because the walk is translation invariant, we may set $g=e$ without loss of generality.

\paragraph{Heuristic solution of the DS problem.} Based on the definitions given above, we can now precisely formulate the DS problem. The question is to identify the value of the iteration $k$ such that, given a small threshold $\epsilon$,
\begin{equation}
    D(T^{*k^*}, U) < \epsilon.
\end{equation}
Physically, if repeated samples of the RW after $k^*(\epsilon)$ steps are taken (always with resetting to the initial state), the resulting distribution is uniform up to a total (small) deviation $\epsilon$. 

The proof of DS leading to Eq.~\eqref{nlogn} is made of two parts: an upper bound, which requires tools of group theory such as Froebenius' formula for characters, and a lower bound, which admits a powerful heuristic explanation in terms of the coupon collector problem, which we now outline -- we refer the reader to Appendix~\ref{a_lower_bound_coupon_collector} for a more technical discussion. 

In the coupon collector problem, one of $n$ distinct coupons is sampled uniformly at each step. The question is how long it takes to observe every coupon at least once. Initially, new coupons are easy to find. Near the end, however, most samples are repetitions. If only one coupon remains unseen, the probability of finding it on the next draw is $1/n$, so the expected waiting time is $n$. More generally, if $j$ coupons remain unseen, the probability of drawing one of them is $j/n$, and the expected waiting time is $n/j$. The total expected collection time is therefore
\begin{equation}
    n+\frac{n}{2}+\frac{n}{3}+\cdots+1
    \sim n\log n.
\end{equation}

After $k$ random-transposition steps, $2k$ labels have been sampled, including repetitions. The coupon collector problem therefore implies that approximately $n\log n$ label selections, or $\frac12 n\log n$ transposition steps, are required to touch every label. Since an untouched label remains in its initial position, this argument heuristically yields the lower-bound scale
\begin{equation}
    k^*_{<}\sim\frac12 n\log n.
\end{equation}

\begin{figure}[t]
\centering
\includegraphics[width=0.75\textwidth]{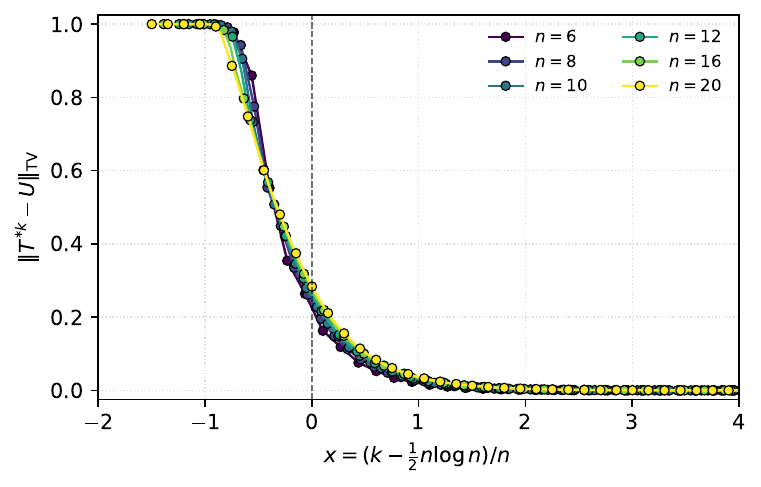}
\caption{Evolution of the measure $\|T^{*k}-U\|_{\mathrm{TV}}$ of the DS random-transposition problem versus the rescaled variable $x = (k - \frac12 n \log n)/n$. A sharp transition occurs around 
$x\simeq 0$. $x<0$ corresponds to an ``ordered phase'', where the RW has memory of the initial condition. On the other hand, the region $x>0$ contains mostly ``random'' configurations, each occurring with almost equal probability.}
\label{sharp}
\end{figure}

This argument identifies the correct timescale. Well before the threshold, the distribution remains strongly nonuniform: many labels have not yet been touched, and the walk retains clear memory of its initial state. Nevertheless, the walk exhibits a genuine cutoff around $k^*$: the transition from order to randomness occurs abruptly within a window of width $O(n)$ (see Fig.~\ref{sharp}). Across this window, the remaining large-scale structure disappears almost at once, in a manner reminiscent of a phase transition.

\paragraph{Cyclic Permutations.}
The DS theorem concerns the random walk on $S_n$ obtained by choosing two labels independently and uniformly. A natural generalization is obtained by defining a probability measure supported on the conjugacy class of $q$-cycles.

The symmetric group decomposes into disjoint conjugacy classes,
\begin{equation}
    S_n=\bigcup_\mu\mathcal C_\mu,
\end{equation}
which are indexed by partitions $\mu=(\mu_1,\mu_2,\ldots)$ of $n$. These partitions are represented by Young diagrams consisting of rows of nonincreasing lengths,
\begin{equation}
    \mu_1\geq\mu_2\geq\ldots,
    \qquad
    \sum_i\mu_i=n.
\end{equation}
For $S_n$, Young diagrams label not only conjugacy classes but also the irreducible representations of the group. We use $\mu$ for conjugacy classes and $\lambda$ for irreducible representations (irreps).

A $q$-cycle has parity $(-1)^{q-1}$. If $q$ is even, each step changes the parity of the current permutation, and the discrete-time walk alternates between the even and odd sectors. If $q$ is odd, each cycle is an even permutation and therefore preserves parity. A walk initialized at the identity then remains in the alternating group $A_n$. Accordingly, for odd $q$, the equilibrium measure must be defined on $A_n$.

The number of $q$-cycles is \cite{sagan2001symmetric}
\begin{equation}
    N_q
    =
    \binom{n}{q}(q-1)!
    \underset{n\gg1}{\sim}
    \frac{n^q}{q}.
\end{equation}
Here, the binomial coefficient counts the possible choices of $q$ labels, while $(q-1)!$ counts their distinct cyclic orderings. A generalization of the DS measure is therefore
\begin{equation}
\label{ppar}
    T(e)=p,
    \qquad
    T(\sigma)=\frac{1-p}{N_q}
\end{equation}
where $p$ is a real positive number less than 1 and 
$\sigma$ is a $q$-cycle. Every other permutation has zero probability.

The coupon collector argument also applies to random $q$-cycles, since each step touches $q$ labels. After $k$ steps, there have been $qk$ label selections, and the scale required to observe every label is determined by
\begin{equation}
    qk\sim n\log n.
\end{equation}
This gives
\begin{equation}
    k_q^*\sim\frac{n}{q}\log n.
\end{equation}
The mixing behavior of random $q$-cycles was established in Ref.~\cite{berestycki2011mixing}.

For the discrete-time walk with even $q$, a nonzero holding probability is required for mixing to the uniform distribution. The reason is that such permutations have odd parity and therefore alternate the walk between the alternating subgroup $A_n$ and its complement. A nonzero holding probability $p$ prevents such alternation from happening. Alternatively, periodicity can be removed by taking the jumps to occur in continuous time according to a Poisson process. We introduce this continuous-time random walk on $S_n$ in the next section.

\section{Classical continuous-time random walk}
\label{s_classical}
The DS random-transposition shuffle provides a standard example of a random walk on the symmetric group $S_n$. We showed that it can be naturally framed as a discrete-time Markov chain: at fixed time intervals the state is applied a random transposition or remains unchanged. Since the ultimate goal of this paper is to address the mixing time of a quantum system evolving in continuous time, we devote this section to casting the discrete-time Markov chain into a classical continuous-time random walk (CCTRW).

In a CCTRW, an elementary move is not applied deterministically at fixed time intervals; rather a transition occurs at exponentially distributed time intervals with a rate $\gamma$. Let $N(t)$ be the number of transitions in an interval $t$. The probability of observing $m$ events is given by
\begin{equation}
    \mathbb{P}[N(t) = m] = \frac{(\gamma t)^m}{m!} e^{- \gamma t}.
\end{equation} 

In analogy to the discrete-time case, the CCTRW is defined by a probability measure $T$ on $S_n$.  The transition operator $H$ is a $n! \times n!$ matrix whose $(\sigma, \sigma')$ element is
\begin{equation}
    H(\sigma\to\sigma') = T(\sigma'\sigma^{-1}). 
\end{equation}
Since $T$ is normalized, $H$ is bistochastic. 

The state space of dimension $n!$ is the regular representation of $S_n$. We denote the basis vectors by $\ket{g}$, with $g\in S_n$. Permutations are not only basis states, but also act as operators on this space. The matrix $R(g)$ is defined by left multiplication as $ R(g)\ket{h} = \ket{g h}$.

Without loss of generality, we are free to set the parameter $p = 0$ by properly rescaling the transition rate $\gamma$: $\gamma \to \gamma' = (1-p)\gamma$. Notice that this freedom is absent in the discrete case, and $p\neq 0$ in eq.\,(\ref{ppar}) is necessary for even-$q$ walks to approach a stationary distribution at late times.

The master equation governing the evolution of the CCTRW is as follows. Let $p(\sigma, t)$ be the probability at time $t$ of the element $\sigma \in S_n$. The equation for $p(\sigma', t + \dd t)$ is
\begin{equation}
    p(\sigma', t + \dd t) = (1 - \gamma \dd t) p(\sigma', t) + \sum_\sigma \gamma \,\dd t \,H(\sigma\to\sigma') \, p(\sigma, t).
\end{equation}
The first term accounts for the permanence of the state in, while the second represents the independent permutations $\sigma\to\sigma'$. 

The master equation becomes a first-order differential equation for the vectorized probability $\bm{p}(t)$
\begin{equation}
    \frac{\dd \bm{p}}{\dd t} = \mathcal{L} \bm{p}, \quad \mathcal{L} = \gamma(H - \bm{1}),
\end{equation}
with initial condition $[\bm{p}(0)]_\sigma = \delta_{\sigma, e}$. The integration is immediate and yields 
\begin{equation}
\label{CCTRW-prob}
    \bm{p}(t) = e^{t\,\mathcal{L}} \bm{p}(0).
\end{equation}

\paragraph{Stationary distribution.} The largest eigenvalue of $H$ is associated to the vector $\ket{[n]} = \frac{1}{\sqrt{n!}} \sum_{g \in S_n} \ket{g}$. This vector is the one-dimensional basis of the symmetric representation $[n]$ of $S_n$ satisfies
\begin{equation}
    H \ket{[n]} = \ket{[n]}
\end{equation}
for any transition matrix. Indeed, this follows from $\ket{[n]}$ being invariant under the action of any $g\in S_n$: $g\ket{[n]} = \ket{[n]}$. Therefore, the stationary distribution is the uniform one
\begin{equation}
    p_{\infty}(g) = \frac{1}{n!}.
\end{equation}

\paragraph{Transition Matrix of $q$-cycles.} 
We focus on transition matrices whose elementary moves contain all elements of a $q$-cycle occurring with equal probability. We let $\mathcal{C}_q = \mathcal{C}_{[q,1^{n-q}]}$ be the conjugacy class and write such operators as
\begin{equation}
\label{eq_transition_matrices_PH_hamiltonian}
    H_q = \frac{1}{N_q}\sum_{g \in \mathcal{C}_q} g.
\end{equation}
The numerical prefactor $N_q$ ensures normalization. This operator can act on different representations rather than the regular one, for instance those for which some objects are indistinguishable -- see e.g. \cite{statisticalsignatures}. When $H_q$ acts on the regular representation of $S_n$, it gets promoted to a matrix\footnote{In this work, with abuse of notation, $H_q$ will denote both the operator and the matrix in the regular representation, as we will not consider other reducible representations of $S_n$.} with $g\to R(g)$. For each row of $H_q$ has $N_q$ nonzero entries with value one. Therefore, the degree of row-sparsity is $N_q/n! \sim n^{q-n}/q \to 0$ as $n\to\infty$. 
The regular representation is reducible into the irreps $\lambda$ of $S_n$, each occurring with multiplicity $m_\lambda = d_\lambda$, where $d_\lambda$ is its dimension (which can be computed, for instance, through the hook formula). It follows that $n! = \sum_\lambda d_\lambda^2$.

Operators such as $H_q$ are called ``class operators'', since they belong to the center of the group algebra. In other words, for any $h \in S_n$, they satisfy $h H_q h^{-1} = H_q$ or equivalently $[h, H_q] = 0$. By Schur's lemma, when projected onto an irrep $\lambda$, the resulting matrix is proportional to the identity, with eigenvalue
\begin{equation}
    \label{eq_spectrum_class}
    E_q(\lambda) = \frac{\chi_q(\lambda)}{d_\lambda},
\end{equation} 
where $\chi_q(\lambda) = \operatorname{Tr} \lambda(g)$ (with $g \in \mathcal{C}_{q}$) is the character of the $q$-cycle class in irrep $\lambda$. Thus, $H_q$ is solvable without exact diagonalization: the spectrum is given by the set of eigenvalues $\{E_q(\lambda)\}$, each with degeneracy $d_\lambda^2$. The characters can be evaluated, for instance, through Frobenius' formula \cite{hamermesh}. 

Remarkably, for such class operators $H_q$ the structure of the overlaps between permutation states can be written explicitly in terms of characters, which for the $S_n$ are real-valued:
\begin{equation}
    \begin{aligned}
        \braket{h| H_q | g} &= \sum_{\mu} \sum_{s \in S_n} E_q(\mu)\frac{d_\mu}{n!} \chi_{\mu}(s) \braket{h|R(s)|g} = \sum_{\mu} E_q(\mu) \frac{d_\mu}{n!} \chi_{\mu}(h g^{-1}).
    \end{aligned}
\end{equation}
In the first sum, $\mu$ denotes a conjugacy class of $S_n$.

The latter equation implies that $\bm{p}(t)$ is constant among elements belonging to the same conjugacy class, i.e.
\begin{equation}
    p_t(\mu) = \sum_{g \in C_\mu} p_t (g) = \frac{N_\mu}{n!} \sum_\lambda d_\lambda \chi_\lambda(\mu) e^{\gamma (E_q(\lambda) - 1) t}\;. 
\end{equation}
Therefore, if we look at the probability of the event $g \in C_\mu$, we observe that at long times, the conjugacy classes mix independently to the uniform distribution 
\begin{equation}
    p_t(\mu) \underset{t\to\infty}{\sim} \frac{N_\mu}{n!}.
\end{equation}

\paragraph{Subleading Eigenvalues.} For the discrete random walk of Section~\ref{s_DS}, the coupon collector problem allows us to identitify a lower bound for the mixing time. As we shall now see, the same time-scale $O(n\log n)$ arises from the analysis of the first subleading eigenvalue of the CCTRW, which lies in the irrep $[n-1,1]$, of dimension $(n-1)$. The character is $\chi_{[n-1,1]}^q = n-q-1$ and therefore $E_{[n-1,1]} = (n-q-1)/(n-1)$. This implies that for any element $g$, we have\footnote{
    The expansion of $p_t(g)$ in irreps is a specialization of the Fourier decomposition on groups, which we now recall. Let $\mu$ be an irrep of dimension $d_\mu$. Then for any function $f: G\to \mathbb{C}$, one defines the Fourier transform
\begin{equation*}
%\label{Fourier}
    \hat{f}(\mu) = \sum_{g \in G} f(g)\mu(g),
\end{equation*}
together with its inverse:
\begin{equation*}
%\label{Fourier-inverse}
    f(g) = \frac{1}{ |G| } \sum_{\mu} d_\mu \operatorname{tr} \left(\hat{f}(\mu) \mu(g^{-1})\right).
\end{equation*}
Notice that while $f$ is a complex scalar function, $\hat{f}(\mu)$ is generally a complex-valued matrix of size $d_\mu$.
}
\begin{equation}
\label{eq_CCTRW_asympt}
    p_t (g) = \frac{1}{n!} + e^{- q \gamma t/(n-1)} \frac{n-1}{n!} \chi_{[n-1,1]}(g) + \ldots 
\end{equation}
It follows that the variation distance is at leading order
\begin{equation}
    D(p_t, U) \sim \frac{1}{2} e^{- q \gamma t/(n-1)} (n-1) \frac{1}{n!}\sum_g | \chi_{[n-1,1]}(g) |.
\end{equation}
We are then reduced to evaluating $\sum_g | \chi_{[n-1,1]}(g) |$.

The character $\chi_{[n-1,1]}(g)$ is computed, for instance, by inspecting the ``natural representation'' $V = [n] \oplus [n-1, 1]$. Indeed, one has $\chi_{V}(g) = 1 + \chi_{[n-1,1]}(g)$. $V$ is $n$-dimensional and the basis states are the \textit{labels} $\ket{j}$. Thus, the action of a permutation element is $P(g)\ket{j} = \ket{g(j)}$. The character $\chi_{V}(g) = \sum_j \braket{j| P(g) |j} = \operatorname{Fix}(g)$ is the number of \textit{untouched} objects (fixed points) by the permutation $g$. Therefore, $\chi_{[n-1, 1]}(g) = \operatorname{Fix}(g) - 1$. 

We now evaluate $\sum_g | \chi_{[n-1,1]}(g) |$. The sum can be split into two: the first is over the set $\mathcal{A}_1 = \set{g|\operatorname{Fix}(g) = 0}$ and the other over $\mathcal{A}_2=\set{g|\operatorname{Fix}(g) \geq 1}$. The first term coincides with the cardinality of $\mathcal{A}_1$, i.e. the number of \textit{derangements} $\mathcal{A}_1 = D_n$ of $n$ objects (i.e. all objects are not found in their original place). Asymptotically, we have that $D_n \sim n!/e$. The second term also evaluates to $D_n$\footnote{To see that $\sum_{g\in \mathcal{A}_2} (\operatorname{Fix}(g)-1) = D_n$, we proceed as follows. The first term is $\sum_g \operatorname{Fix}(g) = n!$, as $\operatorname{Fix}(g) = 0$ does not contribute. The $n!$ comes from choosing a label $j$ and couting the permutations which leave $j$ invariant. There are $(n-1)!$ of them. Thus $\sum_g \operatorname{Fix}(g) = \sum_{j=1}^n (n-1)! = n!$. On the other hand, the second term is $\sum_{g\in \mathcal{A}_2} = n! - D_n$. Overall, $\sum_{g\in \mathcal{A}_2} (\operatorname{Fix}(g)-1) = n! - (n!-D_n) = D_n$.}. Thus $\sum_g | \chi_{[n-1,1]}(g) | = 2 D_n$ and 
\begin{equation}
    D(p_t, U) \sim e^{- q \gamma t/(n-1)} \frac{(n-1)}{n!} D_n.
\end{equation}

Given a tolerance $\epsilon$, the mixing time \cite{aharonov2001quantum,marquezino2008mixing,saloff2004random} is 
\begin{equation}
    t_{\text{mix}}(q) > \frac{(n-1)}{q\gamma} \log \frac{(n-1) D_n}{\epsilon n!} \underset{n\gg1}{\sim} \frac{n}{q \gamma} \log \frac{n}{\epsilon e} ,
\end{equation}
which is consistent with the DS time-scale of the mixing time $k_* = O(n\log n)$.

\paragraph{Cutoff phenomenon.}
As mentioned in Sec.~\ref{s_DS}, the discrete DS random walk features a sharp cutoff phenomenon in crossing the mixing time threshold. The same occurs for the CCTRW with $q$-cycles. We restrict to even values of $q$ (since odd-values mix only within the alternating subgroup $A_n$) and study this phenomenon numerically.  

The theoretical sharp cutoff is $t^{(q)}_{c}=\frac{1}{q\gamma}n\log n$. More precisely, for any $\epsilon>0$ and for sufficiently large $n$, there exists a $c_{\epsilon,q}>0$ such that \cite{berestycki2011mixing, teyssier2020limit,saloff2004random,levin2026markov,fulman2008convergence}
\begin{equation}
	D(p_t,U)< \epsilon \quad \text{at}\quad t_{\text{mix}}^{(q)} (\epsilon,n)= t_c^{(q)}+c_{\epsilon,q}n.
\end{equation}
In the exact numerical analysis, it is convenient to study the variable 
\begin{align}
	x_{\epsilon}^{(q)}(n) = \frac{\gamma q t_{\text{mix}}^{{(q)}}(\epsilon,n)}{n} -\log n,
\end{align}
which, for large enough $n$ approaches the constant value $x_{\epsilon}^{(q)}(n) \rightarrow\gamma q c_{\epsilon,q}= c_{\epsilon}$, where $c_{\epsilon}$ is independent of $q$.

\begin{figure}
	\centering
	\includegraphics[width=1\linewidth]{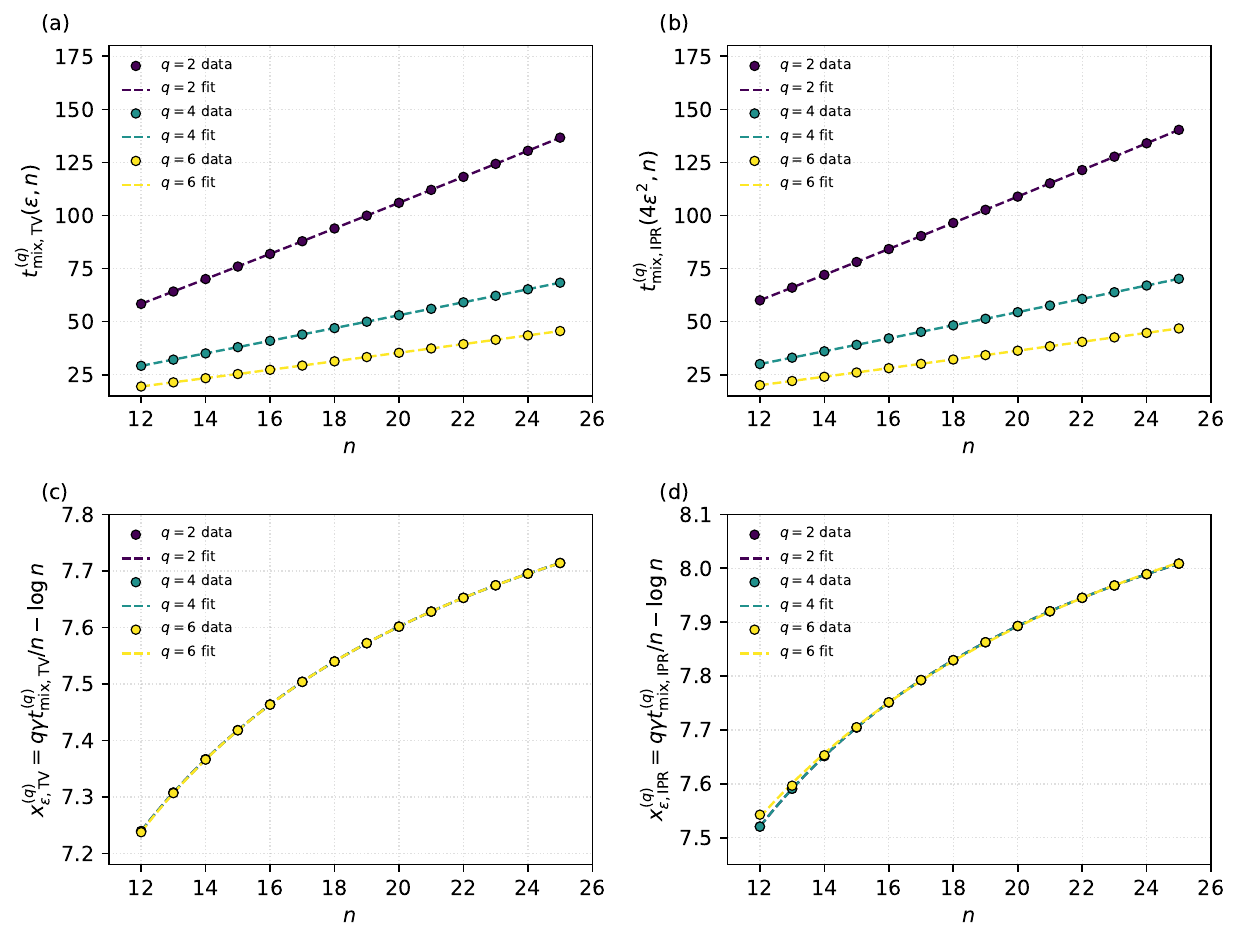}
	\caption{ Mixing times determined from (a) the TV distance for the threshold $\epsilon=10^{-4}$, (b) $L_2$ (or equivalently IPR) for the threshold $4 \epsilon^2$. This choice of thresholds puts the mixing times on the same scale. We study the range $n\in \left[12,25\right]$ and $q = \{2, 4, 6\}$. Numerically extracted mixing times are represented as dots and the dashed lines are the corresponding fit with finite-size corrections up to $O(1/n^2)$. Panels (c) and (d) show the scaling variable $x_{\epsilon}^{(q)}(n) = \frac{\gamma q t_{\text{mix}}^{{(q)}}(\epsilon,n)}{n} \log n -\log n$ for the TV and IPR distances, respectively. Overall, we observe that for the CCTRW, the choice of the TV and IPR distances produces qualitatively similar scalings of the mixing times, hence these measures are operationally equivalent.}
	\label{classical-mixing time-DTV-pt-IRP}
\end{figure}

In Fig.~\ref{classical-mixing time-DTV-pt-IRP} we report our numerical study for $n \lesssim 25$. We observe that the variable $ x_{\epsilon}^{(q)}(n)$ is not a constant for finite $n$, rather it is subject to finite-size effects, which can be captured by polynomial corrections in $1/n$. We found that it is sufficient to include corrections up to order $1/n^2$ to accurately fit $x_{\epsilon}^{(q)}(n)$ in the range $n \in [12,25]$. Panels (a) and (c) are dedicated to the TV distance, and (b) and (d) to the inverse participation ratio (IPR), which we will now discuss.

\subsection{Inverse participation ratio in a classical scenario}

So far, we considered the TV distance for the discrete and continuous-time random walks generated by uniform $q$-class permutations. We analyze here another distance, which will be relevant to the study of unitary quantum dynamics: the Inverse Participation Ratio (IPR), defined as 
\begin{equation}
    \operatorname{IPR}(t) = \sum_{g} \left(p_{t}(g)\right)^2 =  D_2(p_t, U) + \frac{1}{n!}.
\end{equation}
The IPR is a measure of ``spreading'', and coincides with the second moment of $p_t(g)$. 

\begin{figure}[!t]
	\centering
	\includegraphics[width=1\linewidth]{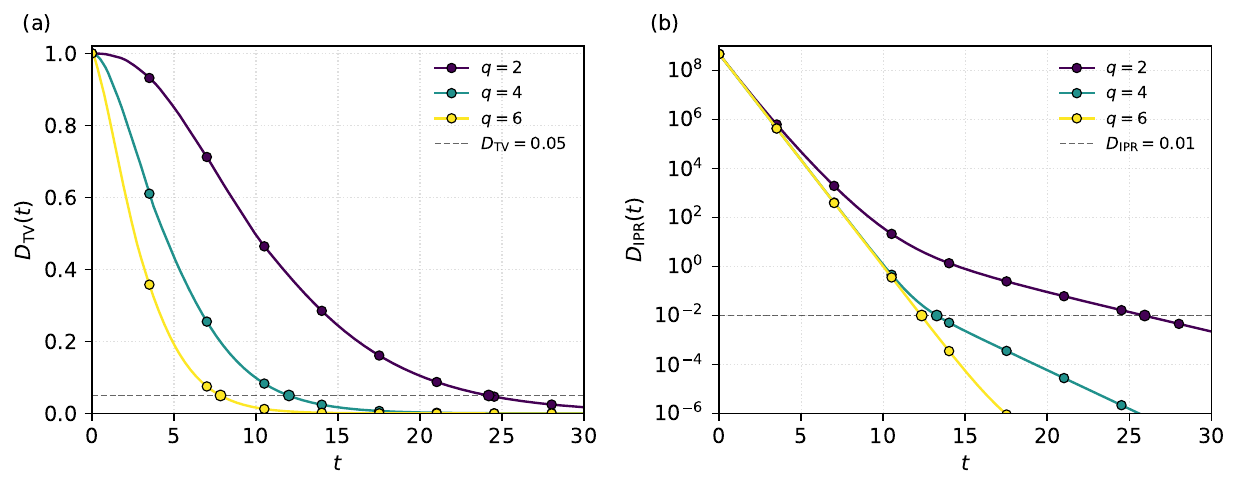}
	\caption{TV distance (a) versus IPR (b) as function of $t$ at $n=12$, where the threshold is set as $\epsilon=0.05$ for (a), $4 \epsilon^2$ for (b). The three curves are associated to CCTRW generated by $H_q$ with $q = \{2,4,6\}$. The mixing times $t^{(q)}_{\mathrm{mix}}$ is marked with vertical dish lines. The larger the cycle $q$, the quicker the decay of the distances.}
	\label{DTV-pt-IRP}
\end{figure}

The IPR inherits the same properties of the TV distance for the CCTRW defined by $H_q$, because it contains only powers of $p_t(g)$ without cross-terms. For instance, conjugacy classes mix independently within the same time-scale $O(n\log n)$: 
\begin{equation}
    \operatorname{IPR}(t, \mu) = \sum_{g \in \mu} \left(p_t(g)\right)^2 \underset{t\to\infty}{\sim} \frac{N_\mu}{(n!)^2}.
\end{equation}
Thus, in the classical setting, there is no qualitative difference in probing mixing through the TV distance or IPR (Fig.~\ref{DTV-pt-IRP}). 

\subsection{Random-Time Measurements and Average Mixing}\label{s_avg_classical}

The TV distances and IPR studied until now are \textit{instantaneous} measures, in the sense that they involve information about measurements at time $t$. In the CCTRW such probability is both the simplest and most effective characterization of the Markov process, because Eq.~\eqref{CCTRW-prob} admits a well-defined stationary distribution. However, when the walk dynamics does not approach a stationary state (as it is the case for the unitary quantum walks studied in the following Section), instantaneous probabilities and distances may oscillate. For this reason, other measures of mixing need to be considered. 

We focus the attention on the one based on uniform measurements on the time interval $[0,t]$. The distribution coincides with the \textit{running average} (RA) of the instantaneous probability
\begin{equation}
\label{eq_running_average}
    \overline{p}_t (g) = \frac{1}{t} \int_0^t \dd t'\; p_{t'} (g),
\end{equation}
with $p_t(g)$ given by Eq.~\eqref{CCTRW-prob}. The limiting distribution as $t\to\infty$ is well-defined and is called the \textit{long-time average} (LTA) $\overline{p}_\infty$. If the stationary distribution of the instantaneous probability is uniform, then the LTA will be uniform too. 

Thus, we can generalize the TV distance Eq.~\eqref{total-variance} to the one involving the RA:
\begin{equation}
    \label{total-variance-RA}
    \overline{D}(P_t, Q) = || \overline{P}_t - Q ||_{\mathrm{TV}} . 
\end{equation}
This distance will define a different mixing time. However, it is related to the instantaneous measure for the CCTRW. To see this, we observe that by convexity
\begin{equation}
\label{eq_convexity_average_bound}
    \overline{D}(p_t, U) \leq \frac{1}{t} \int_0^t \dd t'\; D(p_t, U).
\end{equation}
Let $t$ be a time for which the TV distance $D(p_t, U) < \epsilon$. It then follows that
\begin{equation}
    \overline{D}(p_t, U) \leq \frac{n \log n}{t} + \epsilon \left(1-\frac{n \log n}{t}\right),
\end{equation}
and therefore the expected average mixing time is estimated to be of the order
\begin{equation}
    \overline{t}_{\mathrm{mix}} \sim \frac{n \log n}{\epsilon}.
\end{equation}

\begin{figure}[t]
\centering
\includegraphics[width=0.75\textwidth]{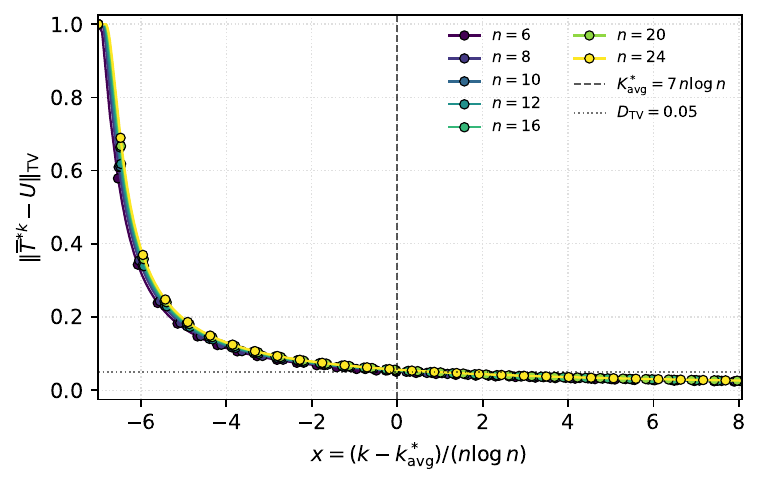}
\caption{Evolution of the average DS measure versus the rescaled variable $x = (k - k^*_{\text{avg}})/n \log n$. While the instantaneous distance features a sharp cut-off phenomenon (see Fig.~\ref{sharp}), the averaged analogue smoothly decrease from its maximal value towards zero. Still, the measure vanishes rapidly. The dotted horizontal line corresponds to the threshold $\epsilon = 0.05$. The value $k^*_{\text{avg}}\approx 7 n \log n$ was chosen to provide a common centering scale. Its magnitude, compared to $\frac{1}{2}n\log n$ of the instantaneous TV distance, shows the broadening effect induced by the running average: it retains and accumulates contributions from small $k$'s, where the distribution is highly non-uniform.}
%\ref{tablech} .) }
\label{sharp-avg}
\end{figure}

Thus, for any fixed threshold $\epsilon$, the instantaneous and time-averaged mixing times share the same leading $O(n \log n)$ dependence on system size, although their subleading corrections differ. In particular, the average induced by uniform measurements removes the sharp cutoff characteristic of the instantaneous distance, as illustrated in Fig.~\ref{sharp-avg}. This estimate is sufficient to the purposes of our work since it implies that we should not expect an average mixing time larger than the instantaneous one for the CCTRW on $S_n$.

%%%%%%%%%%%%%%%%%%%%%%%%%%%%%%%%%%%%%%%%%%%%
\section{Unitary Quantum Walks}
\label{s_quantum}

In Sections~\ref{s_DS} and~\ref{s_classical}, we discussed classical random walks on $S_n$ in discrete and continuous time. The transition matrices $H$, and in particular the class operators defined in Eq.~\eqref{eq_transition_matrices_PH_hamiltonian}, generate walks that, under the appropriate ergodicity conditions, mix toward the uniform distribution on their accessible state space. We now show that the class operators $H_q$ can also be interpreted as permutation Hamiltonians \cite{statisticalsignatures}, generating unitary quantum walks (QWs) on $S_n$.

For a quantum system evolving unitarily, however, one cannot expect irreversible mixing analogous to that of the CCTRW: if  classical mixing involves the gradual loss of information about the initial state, on the contrary unitary evolution preserves the purity of the density matrix. We will return however to the role of information loss in quantum dynamics in the next section.

Despite the intrinsic differences between the CCTRW and the QW generated by the same $H_q$, let's discuss how the QW fails to mix and let's also introduce the diagnostics used to characterize its dynamics. The quantum-mechanical system consists of $n$ sites, each with an on-site Hilbert space of dimension $n$. It contains $n$ particles with $n$ distinct colors, indexed by $i\in\{1,\ldots,n\}$. The total Hilbert space has dimension $n!$, and each basis state $\ket{g}$ corresponds to a permutation $g\in S_n$. The state $\ket{e}=\ket{1\ldots n}$, associated with the identity permutation, is the ``ordered state.'' The computational basis coincides with the state space of the random walk on $S_n$; unlike the classical state space, however, the Hilbert space admits superpositions with complex amplitudes.

The class operators $H_q$ are special cases of Permutation Hamiltonians (PHs). More generally, a PH is a Hermitian combination of group elements,
\begin{equation}
    H = \sum_{g \in S_n} \left(c_g g + c^*_g g^{-1}\right),
\end{equation}
where the $c_g$ are arbitrary coefficients. Hamiltonians of this type are called permutation Hamiltonians because their constituent operators only permute the degrees of freedom at the sites $j\in\{1,\ldots,n\}$. Such PHs commute with every global $SU(n)$ rotation in color space, providing a quantum-mechanical interpretation of their block-diagonal decomposition into irreducible representations of $S_n$\footnote{A PH retains the same functional form when the number of on-site colors is changed; only its matrix representation changes. This makes PHs a versatile class of Hamiltonians in which several physical parameters can be controlled \cite{statisticalsignatures}. For example, ordinary spin chains arise when only two on-site colors are present, and a generic transposition can be written in terms of spin-$1/2$ operators as
\begin{equation*}
    P_{i,j} = 4\; \bm{S}_i \cdot \bm{S}_j + 2\; \bm{1}.
\end{equation*}
If only adjacent transpositions are included, the resulting Hamiltonian
\begin{equation*}
    H_{\text{a}} = J \sum_{j = 1}^n P_{j, j+1}
\end{equation*}
has the same dynamics as the exactly solvable XXX spin chain \cite{Bethe}; its generalization to an arbitrary number of on-site colors is also exactly solvable \cite{Sutherland}. Hamiltonians of the type $H_q$ are class operators and describe all-to-all interactions.}.

We focus on the time evolution of basis states generated by the unitary operator
\begin{equation}
\label{unitary}
    \mathcal{U}(t) = \exp\left(- i H t\right).
\end{equation}
Thus, given an initial state $\ket{\psi(0)}$, its trajectory in Hilbert space is
\begin{equation}
\label{Schroedinger}
|\psi(t)\rangle= \mathcal U(t)|\psi(0)\rangle.
\end{equation}

In contrast to the master equation~\eqref{CCTRW-prob}, which governs the CCTRW probabilities, the Schr\"odinger equation governs probability amplitudes. To facilitate comparison with the CCTRW, we consider measurements in the computational basis. The natural probabilities defining the QW are therefore the transition probabilities between computational-basis states:
\begin{equation}
\label{QW_probability}
    Q(g, h, t) = \left|\braket{h| e^{- i H t} |g}\right|^2.
\end{equation}
Unlike the CCTRW probability distribution in Eq.~\eqref{CCTRW-prob}, $Q(t)$ does not generally approach a stationary distribution at long times. Quantum observables may instead exhibit periodic or quasiperiodic revivals. In particular, class Hamiltonians of the type $H_q$ have rational spectra, so there exists a recurrence time $t_P$ such that $\mathcal{U}(t_P)=\mathcal{U}(0)$; see Appendix~\ref{a_IPR_class}. Because this recurrence time generally grows exponentially with $n$, the thermodynamic limit is often considered when studying the behavior of observables \cite{Sierant_2025}.

Another important issue when comparing CCTRWs and QWs is the normalization of the transition matrix or Hamiltonian. For the CCTRWs of Section~\ref{s_classical}, convergence to the stationary state relies on the normalization of the probability measure entering the transition matrix. By contrast, a QW needs not approach a stationary distribution, so its Hamiltonian needs not be normalized in this way. Nevertheless, using Hamiltonians that coincide with CCTRW transition matrices provides a consistent normalization across different values of $n$ and, for the class operators in Eq.~\eqref{eq_transition_matrices_PH_hamiltonian}, across different values of $q$.

Although the instantaneous distribution $Q$ may have no long-time limit, its RA average $\overline{Q}$, Eq.~\eqref{eq_running_average}, does. Writing the Hamiltonian in spectral form, $H = \sum_n E_n \ket{E_n}\bra{E_n}$, gives the LTA
\begin{equation}
    \overline{Q}_\infty(g) = \sum_E \lvert \braket{g|P_E | e}\rvert^2,
\end{equation}
where $P_E$ is the projector onto the full eigenspace with energy $E$. Thus, the natural quantity for comparing classical and quantum walks on $S_n$ is the averaged distance defined in Eq.~\eqref{total-variance-RA}.

\begin{figure}[t]
\centering
\includegraphics[width=0.75\textwidth]{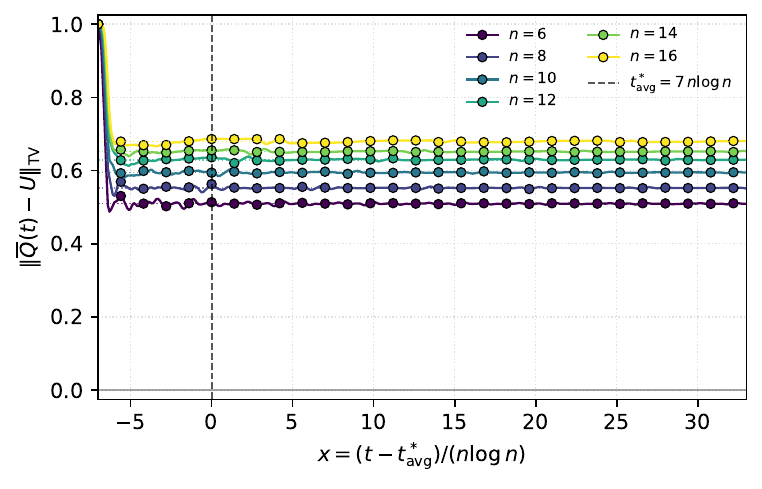}
\caption{Evolution of the average DS measure versus the rescaled variable $x = (T - T^*_{\text{avg}})/n \log n$ for the QW generated by the two-cycle Hamiltonian $H_2$. Contrary to the classical case, the distance of the QW saturates to a finite $n$-dependent value at long times. As in Fig.~\ref{sharp-avg}, the scale $T^*_{\text{avg}} = 7 n \log n$ serves only to separate clearly the initial transient from the late-time behavior of the RA.}
\label{f_average_QW_plateau}
\end{figure}

For the CCTRW, the LTA is the uniform distribution. For a QW, by contrast, the LTA needs not be uniform. For example, for the QW generated by the all-to-all transposition Hamiltonian $H_2$, the distance $\overline{D}$ from the uniform distribution approaches a finite value at long times; see Fig.~\ref{f_average_QW_plateau}. Therefore, one commonly defines the \textit{quantum mixing time} by the condition
\begin{equation}
\label{eq_quantum_mixing_time_distance}
    \overline{D}(Q_t, \overline{Q}_\infty) \leq \epsilon.
\end{equation}
For instance, the average mixing time of a QW on a dense random graph with $N_n$ nodes can be shown to scale as $O(N_n^{3/2})$ \cite{chakraborty2020fast}.

Since $\overline{Q}_\infty(g)\neq 1/n!$ in general, the CCTRW and quantum mixing times cannot be compared directly. Nevertheless, unitary dynamics explores Hilbert space through dephasing and the progressive spreading of probability over an increasing number of computational-basis states. To characterize this process, we introduce the quantum dephasing time as a measure of the onset of coherent quantum spreading.

\subsection{Quantum Dephasing Time}

Although Eq.~\eqref{eq_quantum_mixing_time_distance} provides a measure of mixing under unitary evolution, one can also study higher moments of the probability distribution. The first nontrivial moment is the IPR,
\begin{equation}
    \mathrm{IPR}_{\psi}(t) = \sum_{h \in S_n} \left| \braket{h| e^{- i H t} | \psi} \right|^4,
\end{equation}
where $\ket{\psi}$ is the initial state.

In quantum many-body physics, the IPR is a standard basis-dependent diagnostic: it quantifies the spreading of a state over a given basis \cite{Kramer_1993,misguich2016inverse,tsukerman2017inverse,evers2008anderson,frey2024probing} and is particularly useful in studies of random interactions and localization.
If $\mathrm{IPR}(t)=1$, the probability is concentrated on a single basis state, whereas $\mathrm{IPR}(t)=1/n!$ implies a uniform distribution. Quantum dynamics generally prevents such complete uniform spreading: the long-time average of the IPR is typically greater than $1/n!$.

Like the transition probability $Q$, the IPR generally does not converge as $t\to\infty$ and therefore does not, by itself, provide a common mixing diagnostic for the QW and CCTRW. Nevertheless, its short-time expansion for the ordered initial state is
\begin{equation}
    \operatorname{IPR}(t) \sim 1 - 2\operatorname{Var}_e[H] t^2 + O(t^4),
\end{equation}
where
\begin{equation}
    \operatorname{Var}_e [H] = \braket{e|H^2 |e} - \left(\braket{e|H|e}\right)^2 = \sum_n \lvert \braket{E_n|e}\rvert^2 (E_n^2 - E_n \braket{H}_e).
\end{equation}
Thus, the IPR initially departs from unity quadratically, with a coefficient proportional to the variance of $H$ in the initial state.

We define the purely quantum \textit{dephasing time} by
\begin{equation}
    t_{\text{deph}}\sim \frac{1}{\sqrt{\operatorname{Var}_e [H]}}.
\end{equation}
For $t\ll t_{\text{deph}}$, the IPR remains close to unity, indicating that the state remains concentrated on its initial computational-basis configuration. For $t\gtrsim t_{\text{deph}}$, the IPR departs substantially from unity as the state spreads over the computational basis. This timescale provides a simple, computationally accessible measure of the onset of quantum spreading.

For PHs, the computational basis variance $\operatorname{Var}_e[H]$ is exactly equal to the spectral variance
\begin{equation}
\label{eq_spectral_variance}
    \sigma_E^2 = \int \dd E\; \rho(E) E^2 - \left(\int \dd E\; \rho(E) E\right)^2,
\end{equation}
where $\rho(E) = D_n^{-1} \sum_j \delta(E-E_j)$ is the empirical density of states. Indeed, for every PH, the only matrix element connecting $g$ to itself is the identity permutation, which has coefficient $c_e$. Hence
\begin{equation}
    \operatorname{Var}_e [H] = \sum_{g \neq e} \lvert c_g \rvert^2.
\end{equation}
In the regular representation, $\operatorname{Tr} R(g) = n! \delta_{g,e}$, and therefore $\operatorname{Tr} H = n! c_e$ and $\operatorname{Tr} H^2 = n! \sum_g \lvert c_g \rvert^2$. It follows that
\begin{equation}
    \sigma_E^2 = \frac{1}{n!} \operatorname{Tr} H^2 - \left(\frac{1}{n!}\operatorname{Tr} H\right)^2 = \operatorname{Var}_e [H].
\end{equation}

In particular, for $q$-class PHs, the dephasing time is
\begin{equation}\label{eq_QW_dephasing}
    t_{\text{deph},q} = \sqrt{N_q}\underset{n \gg 1}{\sim} \sqrt{\frac{n^q}{q}}.
\end{equation}
Thus, the IPR leaves its initial plateau on a timescale $O(n^{q/2})$. In Appendix~\ref{a_IPR_class}, we show that the same timescale follows from the statistical behavior of characters in the bulk spectrum. We stress that the dephasing time is a \textit{local} scale: it probes only the first layer of states connected to the identity by a single application of $H$. It therefore contains no information about the exploration of the full Hilbert space, whose computational basis consists of the elements of $S_n$.

\begin{figure}
    \centering
    \subfloat[]{\includegraphics[width=0.45\linewidth]{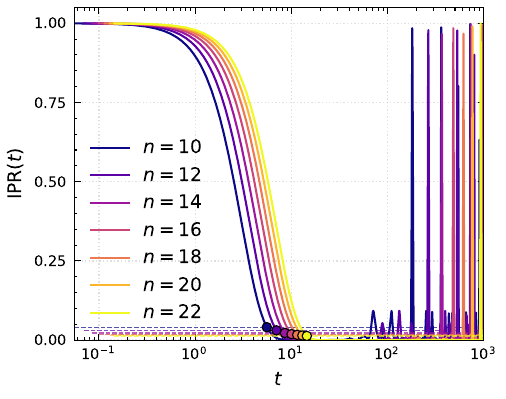}}
    \hfill
    \subfloat[]{\includegraphics[width=0.45\linewidth]{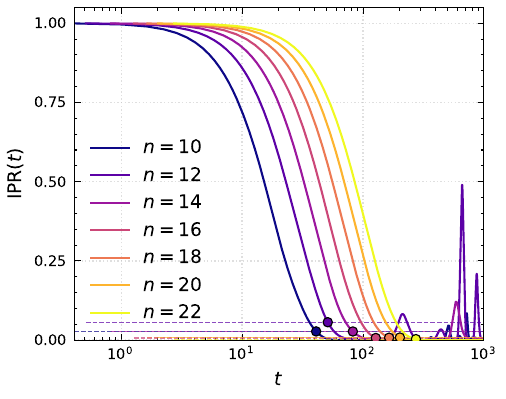}}
    \hfill
    \subfloat[]{\includegraphics[width=0.45\linewidth]{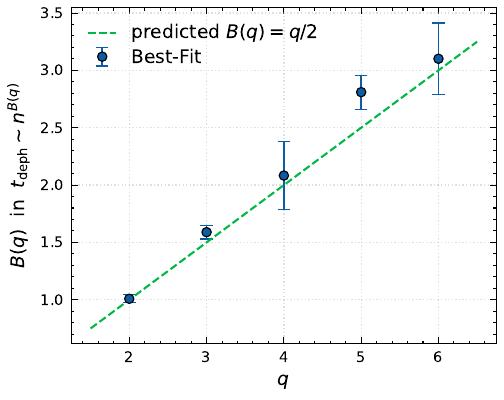}}
    \caption{IPRs of class operators for $n\in[10,23]$. (a) IPR of the transposition Hamiltonian for several values of $n$. The dots mark the quantum dephasing time, estimated from the first crossing with the long-time average. (b) IPR of the 4-cycle class Hamiltonian. As $n$ or $q$ increases, the IPR departs more slowly from unity and characteristic revivals occur at later times. (c) Fitted exponent $B(q)$ in $t_{\text{deph}}(n,q)\sim n^{B(q)}$ as a function of $q$. In contrast to the CCTRW result, $B(q)$ increases with $q$ and agrees with the theoretical prediction $B(q)\sim q/2$.}
    \label{f_IPR2}
\end{figure}

Figure~\ref{f_IPR2} shows the IPR for different $H_q$ and system sizes. As $n$ increases, the initial plateau persists longer and the subsequent decay is delayed. In other words, larger systems exhibit slower dephasing because of spectral crowding. Numerically, we estimate this timescale from the first crossing of the IPR with its LTA, shown by the dashed lines. The resulting timescales exhibit the same scaling as $t_{\text{deph},q}$. This crossing does not imply that the walk has explored a macroscopic fraction of the $n!$ configurations; it identifies only the first time at which the spectral phases have dephased sufficiently for the IPR to cross its LTA.

\section{Quantum Stochastic Walks}\label{s_quantum_stochastic}

The main obstruction in applying the definition of classical mixing to the QW is due to the reversible nature of the evolution: it is not guaranteed that the uniform distribution is reached at long times, either instantaneously or in the average. The natural understanding of this phenomenon comes from information theory: a CCTRW eventually loses information, reaching a distribution of maximum entropy -- the uniform distribution, -- while unitary evolution is reversible and therefore the von Neumann entropy associated to the density matrix $\rho(t) = \ket{\psi(t)}\bra{\psi(t)}$ is zero at all times.

Thus, we consider a protocol, within the theory of quantum stochastic walks \cite{PhysRevA.81.022323}, which satisfies by construction the requirement that the asymptotic probability is uniform. 

We first show how the CCTRW with Poisson rate $\gamma$ can be naturally cast in the form of an open quantum system (OQS) process. To this aim, we extend the states considered from pure states to statistical mixtures generated by density matrices $\rho$. A general density matrix can be written as $\rho=\sum_{s,g}\rho_{s g}(t)\ket{s}\bra{g}$, with $\sum_s\rho_{ss}(t)=1$. Classical probability distributions correspond to the $n!$-dimensional subspace of diagonal density matrices, $\rho=\sum_g p_t(g)\ket{g}\bra{g}$. The 
basis states indexed by group elements transform under the regular representation $R$ of $S_n$ whereas the diagonal density matrices form the corresponding classical probability simplex. The maximally mixed state (MMS) is 
\begin{equation}
    \rho_{\text{MMS}} = \frac{1}{n!} \sum_g \ket{g}\bra{g} = \frac{\bm{1}}{n!}. 
\end{equation}

To understand the emergence of classical dynamics, we can observe that the CCTRW over $q$-cycles is mathematically equivalent to the following OQS setup. Consider the initial state $\rho_0 = \ket{e}\bra{e}$ and study the following Poisson process: after a waiting time $\Delta t \sim \operatorname{Exp}(\gamma)$ the transition
\begin{equation}
    \rho_{j + 1} = L_g \rho_j L_g^\dagger, \quad L_g = \sum_{s \in S_n} \ket{g s} \bra{s} = R(g),
\end{equation}
occurs, where $g$ is randomly chosen from the conjugacy class of $q$-cycles $\mathcal{C}_q$. The operator $L_g(\cdot)L_g^\dagger$ maps $\ket{h}\bra{h}$ to $\ket{g h}\bra{g h}$.

For infinitesimal times $\dd t$, we have the first-order equation
\begin{equation}
    \rho(t + \dd t) = (1- \gamma \dd t) \rho(t) + \frac{\gamma}{N_q} \dd t \sum_{s \in \mathcal{C}_q} L_s \rho(t) L_s^\dagger, 
\end{equation}
which corresponds to the Lindblad equation:
\begin{equation}
\label{eq_limit_CCTRW_lindblad}
    \frac{\dd \rho}{\dd t} = \gamma\left(\Phi_q(\rho) - \rho\right), \qquad \Phi_q(\rho) = \frac{1}{N_q} \sum_{s \in \mathcal{C}_q} R(s) \rho R(s)^\dagger.
\end{equation}
We recall that $p_t(g) = \operatorname{Tr}(\rho(t) \ket{g}\bra{g})$ and therefore the corresponding equation for the probability is
\begin{equation}
\begin{aligned}
    p_{t + \dd t}(g) &= (1- \gamma \dd t) p_t(g) + \frac{\gamma}{N_q} \dd t \sum_{s \in \mathcal{C}_q} \operatorname{Tr}\left(\rho(t) L_s^\dagger \ket{g}\bra{g}L_s\right) \\
    & = (1- \gamma \dd t) p_t(g) + \frac{\gamma}{N_q} \dd t \sum_{s \in \mathcal{C}_q} p_t(s^{-1} g).
\end{aligned}
\end{equation}
This equation coincides with Eq.~\eqref{CCTRW-prob} when the transition matrix is $H_q$. This shows how the CCTRW is retrieved as a particular limit of open quantum dynamics. By appealing to the results of Section~\ref{s_classical}, the stationary state is $\rho_\infty = \rho_{\text{MMS}}$.   

Eq.~\eqref{eq_limit_CCTRW_lindblad} describes an essentially classical process, where no superpositions are created at any time. We can create quantum superpositions by considering evolutions with another $m$-class operator
\begin{equation}
\label{eq_OQS_master_equation}
    \frac{\dd \rho}{\dd t} = - i \alpha [H_m, \rho] + \gamma\left(\Phi_q(\rho) - \rho\right), \qquad \rho_0 = \ket{e}\bra{e}.
\end{equation}
Denoting $\mathcal{Q} = \alpha/\gamma$ the dimensionless  strength, when $\mathcal{Q} \gg 1$, the dynamics is dominated by quantum evolution and approaches the QW on timescales over which dissipation is negligible. When $\mathcal{Q}\to 0$, the dynamics is that of a CCTRW.
From an open-system perspective, the dissipative term in Eq.(\ref{eq_OQS_master_equation}) describes stochastic permutation noise: at Poisson-distributed times, the environment induces a randomly selected \(q\)-cycle among the degrees of freedom. The coherent term implements the corresponding permutation dynamics at the Hamiltonian level. Eq.~\eqref{eq_OQS_master_equation} may therefore be viewed as a minimal setting in which coherent permutation dynamics competes with incoherent permutation events generated by the environment.

The OQS dynamics described by Eq.~\eqref{eq_OQS_master_equation} possesses several properties of physical and mathematical relevance, which allow us to write its exact solution. The first is the fact that $\rho_{\alpha, \gamma}(t)$ lives in the regular subspace and, moreover, the coherent and dissipative superoperators commute. Because $H_m$ is a class operator, it commutes with every $R(s)$ appearing in $\Phi_q$. Consequently, the coherent and dissipative superoperators commute, and the solution factorizes as shown below. This property is not generic for permutation Hamiltonians: for a noncentral Hamiltonian \(H\), one generally has \([H,R(s)]\neq0\), the factorization below no longer follows, and neither the \(\alpha\)-independence of the trace distance nor the bound on the computational-basis mixing established below is guaranteed.

The probabilities of elements belonging to the same conjugacy class are the same at any time, i.e.
\begin{equation}
    p_{\alpha, \gamma}(h g h^{-1}, t) = p_{\alpha, \gamma}(g, t).
\end{equation}
The second property is related to the \textit{trace distance} of $\rho_{\alpha, \gamma}(t)$ at long times, defined as
\begin{equation}
    T(\rho, \sigma) = \frac{1}{2} ||\rho - \sigma||_1 = \frac{1}{2} \operatorname{Tr}\left[\sqrt{(\rho-\sigma)^\dagger(\rho - \sigma)}\right].
\end{equation}
This distance measures the distinguishability of any two density matrices. We first observe that since 
\begin{equation}
    \rho_{\gamma}(t) = e^{\mathscr{L} t} [\rho(0)], \qquad \mathscr{L}[\rho] = \gamma (\Phi(\rho) - \rho),
\end{equation}
is the formal solution to the CCTRW ($\alpha = 0$) limit Eq.~\eqref{eq_limit_CCTRW_lindblad}, the solution of the OQS dynamics Eq.~\eqref{eq_OQS_master_equation} is 
\begin{equation}
    \rho_{\alpha, \gamma}(t) = \mathcal{U}_{\alpha}(t) \rho_{\gamma}(t) \mathcal{U}_{\alpha}^\dagger(t), \qquad \mathcal{U}_\alpha (t) = e^{-i \alpha H_m t}.
\end{equation}
Using the invariance under unitary transformations of the trace distance, we therefore have that
\begin{equation}
\label{eq_trace_invariance}
    T\left(\rho_{\alpha, \gamma}(t), \frac{\bm{1}}{n!}\right) = T\left(\rho_{\gamma}(t), \frac{\bm{1}}{n!}\right),
\end{equation}
or, in other words, the trace distance is independent of $\alpha$. Thus, the addition of superpositions (through $H$) in Eq.~\eqref{eq_OQS_master_equation} does not affect the convergence of $\rho$ to the uniform distribution.

What about the TV distance of the probability distributions? The probability distribution generated by Eq.~\eqref{eq_OQS_master_equation} is
\begin{equation}
\label{eq_proba_quantum_stochastic_walk}
    p_{\alpha, \gamma}(g,t) = \sum_{h \in S_n} Q(g,h,\alpha t) p_\gamma(h,t),
\end{equation}to
where $p_\gamma(h,t)$ is given by Eq.~\eqref{CCTRW-prob} and $Q_{\alpha, m}(g,h,t)$ is the transition probability of the QW Eq.~\eqref{QW_probability}. This result is therefore a convolution of two separate probabilities, one associated with the CCTRW of $q$-cycles and the one of a QW of $m$-cycles. Let us denote
\begin{equation}
    \Delta(\rho) = \sum_g \ket{g} \braket{g|\rho|g} \bra{g} = \sum_g p(g) \ket{g}\bra{g}
\end{equation}
the dephasing map on the computational basis. We can write the TV measure Eq.~\eqref{total-variance} as
\begin{equation}
    D(\alpha, t) = T\left(\Delta(\rho_{\alpha, \gamma}(t)), \Delta \left(\frac{\bm{1}}{n!}\right)\right).
\end{equation}
$\Delta$ projects density matrices in the diagonal ($\ket{g}\bra{g}$) subspace by eliminating cross terms ($\ket{s}\bra{g}$). It is therefore a completely positive trace-preserving map, whose contractivity property yields the inequality 
\begin{equation}
    T\left(\Delta(\rho_{\alpha, \gamma}(t)), \Delta \left(\frac{\bm{1}}{n!}\right)\right) \leq T\left( \rho_{\alpha, \gamma}(t), \frac{\bm{1}}{n!} \right).
\end{equation}
From the trace-invariance, Eq.~\eqref{eq_trace_invariance}, we observe that the r.h.s. of the inequality is the TV measure of the CCTRW, $D(0, t)$, and therefore we arrive at the central inequality of this work
\begin{equation}
\label{eq_bound_alpha}
    D(\alpha, t) \leq D(0, t),
\end{equation}
Thus, although coherent evolution leaves the distance of the full density matrix from the maximally mixed state unchanged, it can only bring the probability distribution obtained in the computational basis closer to uniformity.

It is important to emphasize the operational and basis-dependent meaning of this result. The distance \(D(\alpha,t)\) does not quantify convergence of the full quantum state toward equilibrium, but rather the randomization of the probability distribution obtained by measurements in the computational basis \(\{|g\rangle\}_{g\in S_n}\). This basis is physically distinguished in the present problem because its states correspond precisely to the permutations, forming the configuration space of the classical random walk. In a different measurement basis, the corresponding probability distribution may not satisfy an analogous inequality. The quantum speedup established by Eq.~\eqref{eq_bound_alpha} should therefore be understood as an acceleration of classical output randomization in the permutation basis, rather than as a faster equilibration of the quantum state itself.

As we shall see from the numerical analysis in the next sections, the inequality Eq.~\eqref{eq_bound_alpha} does not appear to be saturated by computational-basis initial states $\rho(0) = \ket{e}\bra{e}$ evolved with $(m, q)$-cycles. In contrast, insensitivity of the mixing time to the coherent coupling strength has been shown to hold for a class of initial states in open quadratic fermionic systems \cite{SciPostPhys.9.4.049}.

\paragraph{Mode decomposition.} The convolution structure of the total probability
\begin{equation}
\label{eq_convolution_kernel}
    p_{\alpha, \gamma}(t) = Q(\alpha t) * p_{\gamma}(t),
\end{equation}
and the constancy over conjugacy classes make the estimation of the Fourier transform immediate. Combining the notions of Section~\ref{s_classical} and Section~\ref{s_quantum} we have that the group Fourier transform of $p_{\alpha, \gamma}(t)$ reads
\begin{equation}
\label{eq_FT_probability}
    \hat{p}_{\alpha, \gamma}(\lambda, t) = \kappa_\lambda (\alpha t) e^{- \gamma(1 - E_q(\lambda)) t} \bm{1}_{d_\lambda}, 
\end{equation}
where $\kappa_\lambda (\alpha t)$ is the Fourier transform of the QW kernel, given by 
\begin{equation}
    \kappa_\lambda(\alpha t) = \frac{1}{d_\lambda} \sum_g \chi_\lambda(g) Q(g, e, \alpha t), \qquad Q(g, e, \alpha t) = \left\lvert \frac{1}{n!} \sum_\lambda d_\lambda e^{- i \alpha t E_m(\lambda)} \chi_\lambda (g)\right\rvert^2. 
\end{equation}

\paragraph{Small $\alpha$ limit.} When $\alpha \ll \gamma$, we can perturbatively expand the evolution of the density matrix. With $H_m = \sum_{s \in \mathcal{C}_m} R(s)/N_m$, we have that the probability matrix deviates quadratically from the identity as
\begin{equation}
    Q_{\alpha t} = \bm{1} + \frac{(\alpha t)^2}{N_m}(H_m - \bm{1}) + O((\alpha t)^4).
\end{equation}
Since probabilities are squared amplitudes, the linear term vanishes and the leading correction of probability $p_{\alpha, \gamma}(t)$ is quadratic in ($\alpha t$) 
\begin{equation}
    p_{\alpha, \gamma} \underset{\alpha \ll \gamma}{\sim} p_{\gamma}(t) + \frac{(\alpha t)^2}{N_m}(H_m - \bm{1}) p_{\gamma}(t) + O((\alpha t)^4).
\end{equation}
At leading order, the unitary evolution acts on the probabilities as an additional \textit{classical} $m$-cycle step. In Fourier space, the quantum kernel reads
\begin{equation}
    \kappa_\lambda(\alpha t) = 1 - \frac{(\alpha t)^2}{N_m}(1-E_m(\lambda) ) + O((\alpha t)^4),
\end{equation}
and therefore the irrep-resolved mode for the full probability is
\begin{equation}
    \hat{p}_{\alpha, \gamma}(\lambda, t)\underset{\alpha \ll \gamma}{\sim} e^{- \gamma (1 - E_q (\lambda)) t} \left(1 - \frac{(\alpha t)^2}{N_m}(1-E_m(\lambda) )\right) + O((\alpha t)^4) .
\end{equation}

\subsection{The case of $S_2$}

$S_2$ is a group of cardinality $2$ and therefore one can use the basis of Pauli operators to resolve the dynamics of the OQS. There is a single transposition element $\tau = P_{1,2}$. Denote Pauli matrices as
\begin{equation}
    X = \begin{bmatrix}
        0 & 1 \\
        1 & 0
    \end{bmatrix}, \quad Y = \begin{bmatrix}
        0 & -i \\
        i & 0
    \end{bmatrix},\quad
    Z = \begin{bmatrix}
        1 & 0 \\
        0 & -1
    \end{bmatrix}.
\end{equation}
Clearly, $X$ is bistochastic and defines the regular representation of $\tau$. We therefore take
\begin{equation}
    H = R(\tau) = X, \quad \Phi(\rho) = X \rho X.
\end{equation}
Starting from the identity $\rho_0 = \ket{e}\bra{e}$, the evolution is exact since $X^2 = \bm{1}$:
\begin{equation}
    \rho_{\alpha, \gamma}(t) = \frac{1}{2} \left[\bm{1} + e^{-2\gamma t}\left(\cos(2\alpha t) Z + \sin(2\alpha t) Y\right)\right].
\end{equation}
Its trace-distance follows exactly the exponential decay
\begin{equation}
    T\left(\rho_{\alpha, \gamma}(t) , \frac{\bm{1}}{2}\right) = \frac{1}{2}e^{- 2 \gamma t},
\end{equation}
independent of $\alpha$. 
\iffalse
The mixing time therefore is 
\begin{equation}
    t_{\operatorname{mix}}(\gamma, \epsilon) = \frac{1}{2\gamma} \left(\log \frac{1}{2\epsilon} \right), \quad \epsilon < \frac{1}{2}.
\end{equation}
\fi
Similarly, the probabilities are
\begin{equation}
    p_{\alpha, \gamma}(e,t) = \frac{1}{2} + \frac{1}{2}e^{-2\gamma t}\cos(2\alpha t), \qquad  p_{\alpha, \gamma}(\tau, t) = \frac{1}{2} - \frac{1}{2}e^{-2\gamma t}\cos(2\alpha t)
\end{equation}
and the TV distance reads
\begin{equation}
    D(\alpha, t) = \frac{1}{2}e^{- 2 \gamma t} | \cos(2\alpha t) |.
\end{equation}
Alike the unitary QW dynamics discussed in Section~\ref{s_quantum}, the TV distance features an oscillatory factor. Such oscillations are due to the competition of two modes (irreps) for $S_2$. As shown numerically below, for larger $n$ the residual oscillations become relevant mainly at very small values of $\epsilon$ or when $\alpha \gg \gamma$.

From this minimal example we observe the mechanism at play: while the trace distance remains $\alpha$-independent, the coherences generated by quantum dynamics, here in the nonzero elements $\ket{e}\bra{\tau}$ and $\ket{\tau}\bra{e}$, allow the information to be distributed on a larger space.

Thus, the case of $S_2$ is the minimal example in which the addition of quantum coherences accelerates mixing in the computational basis. 

\subsection{Dynamics of the Slowest Mode}

Because of the damping factor in Eq.~\eqref{eq_FT_probability}, for sufficiently large times, the irrep $[n-1,1]$ is the slowest one to decay. In analogy to Section~\ref{s_classical}, we discuss the contribution of this irrep to the mixing time. While Eq.~\eqref{eq_bound_alpha} may be impractical to study analytically, we provide here an analysis of the slow-mode estimate. Let us begin from the expansion of the total probability
\begin{equation}
    p_{\alpha, \gamma}(g,t) = \frac{1}{n!} \sum_{\lambda} d_\lambda \hat{p}_{\alpha, \gamma}(\lambda, t) \chi_\lambda(g) = \frac{1}{n!} + \frac{n-1}{n!} \hat{p}_{\alpha, \gamma}([n-1,1], t)\chi_{[n-1,1]}(g) + \ldots \;.
\end{equation}
In the purely classical case $\alpha = 0$, the analysis of such irrep is sufficient to inform about the mixing timescale -- see Section~\ref{s_classical}. We therefore aim at computing the QW kernel, which coincides with the ratio 
\begin{equation}
    \frac{\hat{p}_{\alpha, \gamma}([n-1,1], t)}{\hat{p}_{0, \gamma}([n-1,1], t)}=\kappa_{[n-1,1]}(\alpha t),
\end{equation}
to characterize the dynamics of the slowest mode. 

Using the definition of Kronecker coefficient associated to irreps $(\lambda, \rho, \sigma)$
\begin{equation}
    g_{\lambda, \rho, \sigma} = \frac{1}{n!} \sum_{g \in S_n} \chi_{\lambda}(g) \chi_{\rho}(g) \chi_{\sigma}(g),
\end{equation}
we obtain the following expression for $\kappa_{[n-1,1]}(\alpha t)$:
\begin{equation}
    \kappa_{[n-1,1]}(\alpha t) = \sum_{\rho, \sigma} \frac{g_{[n-1,1], \rho, \sigma} d_\rho d_\sigma}{n! (n-1)} \cos( X_m(\rho, \sigma) \alpha t),\quad X_m(\rho, \sigma) = E_m(\rho) - E_m(\sigma). 
\end{equation}
In contrast to the classical case, the case $\alpha \neq 0$ involves mixing of irreps. 

Nonetheless, the Kronecker coefficient can be given a probabilistic interpretation. From the identity
\begin{equation}
    \sum_{\rho, \sigma} d_\rho d_\sigma g_{\lambda, \rho, \sigma} = n! d_\lambda,
\end{equation}
we observe that the weight
\begin{equation}
    w_{\rho, \sigma}^\lambda = \frac{d_\rho d_\sigma g_{\lambda, \rho, \sigma}}{n! d_\lambda}\leq 1, 
\end{equation}
is such that $\sum_{\rho, \sigma}  w_{\rho, \sigma}^\lambda = 1$ and can therefore be interpreted as a probability measure on the joint space of two irreps. The marginals w.r.t. a single irrep yield the Plancherel measure $\sum_\sigma w_{\rho, \sigma}^\lambda = d_\rho^2/n!$. 
This reformulation converts the quantum suppression of the slowest classical mode into a problem about the statistics of energy differences under a well-defined probability measure on pairs of irreps.

Therefore, the slow-mode ratio becomes
\begin{equation}
    \kappa_{\lambda_S}(\alpha t) = \mathbb{E}_S [e^{i \alpha t X_m}] = \mathbb{E}_{S} [\cos(\alpha t X_m)]
\end{equation}
and thus coincides with the characteristic function of $X_m(\rho, \sigma)$. 

For nontrivial cycles, we have the following exact identities, which follow from character multiplication and column orthogonality
\begin{equation}
    \mathbb{E}_{S}[E_m(\rho)] = 0, \qquad \mathbb{E}_S[E_m(\rho)^2] = \frac{1}{N_m}, \qquad \mathbb{E}_S[E_m(\rho)E_m(\sigma)] = \frac{E_m(\lambda_S)}{N_m}. 
\end{equation}
Since $E_m(\lambda_S) = (n - m - 1)/(n-1)$, we obtain that the variance of $X_m$ is
\begin{equation}
    \operatorname{Var}_S [X_m] =2 \left[\frac{1}{N_m}- \frac{E_m(\lambda_
    S)}{N_m}\right]= \frac{2m}{(n-1) N_m} \underset{n \gg 1}{\sim} \frac{2 m^2}{n^{m+1}}.
\end{equation}
This allows us to compute the statistical correlation of any two energies
\begin{equation}
    \operatorname{Corr}_S[E_m(\rho), E_m(\sigma)] = \frac{\operatorname{Cov}_S[E_m(\rho), E_m(\sigma)]}{\sqrt{\mathbb{E}_S[E_m(\rho)^2]\mathbb{E}_S[E_m(\sigma)^2]}} = 1 - \frac{m}{n-1},
\end{equation}
which tends to unity (perfect correlation) as $n$ increases.

Using the relation $\cos(x) \geq 1 - x^2/2$, we have that
\begin{equation}
\label{eq_bound_kappa_s}
    \kappa_S (\alpha t) \geq 1 - \frac{(\alpha t)^2}{2} \operatorname{Var}_S[X_m] = 1 - \frac{m (\alpha t)^2}{(n-1) N_m} \underset{n \gg 1}{\sim} 1- \frac{m^2}{n^{m+1}} \alpha^2 t^2. 
\end{equation}
If we define the quantum slow dephasing time as
\begin{equation}
    t_{S}(\alpha) = \frac{n^{\frac{m+1}{2}}}{\alpha m},
\end{equation}
then the bound Eq.~\eqref{eq_bound_kappa_s} takes the form
\begin{equation}
     \kappa_S (\alpha t) \geq 1 - \left(\frac{t}{t_S(\alpha)}\right)^2.
\end{equation}

While the analysis of $\kappa_S(\alpha t)$ is not by itself sufficient to extract a precise ratio 
\begin{equation}
    r(\alpha, \gamma,t ) = \frac{D(\alpha, t)}{D(0, t)},
\end{equation}
we observe that coherent dynamics can not substantially suppress the slow mode before the quantum slow dephasing time unless $t \sim t_S(\alpha)$. This suggests a necessary criterion for the impact of quantum coherence on classical mixing. The classical slow-mode mixing occurs in a window of time $t^{(0)}_{\text{mix}, S} = \frac{n}{\gamma q} \log \frac{n}{e \epsilon}$. Therefore, equating $t^{(0)}_{\text{mix}, S} = t_S(\alpha)$ gives an estimate for the `critical' value of the dimensionless ratio $\mathcal{Q} = \alpha/\gamma$ where quantum coherences generate an $O(1)$ modification of the slow mode from the classical mixing time:
\begin{equation}\label{eq_Q_star}
    \mathcal{Q}_{*} \sim \frac{q}{m} \frac{n^{(m-1)/2}}{\log\frac{n}{e \epsilon}}. 
\end{equation}
For example, in the case of random transpositions, one would need to set $\mathcal{Q}_* \sim \sqrt{n}/\log \frac{n}{e \epsilon}$. 

\subsection{Numerical Results}\label{sub_numerics}

We now investigate numerically the computation of the quantum-to-classical ratio $r(\alpha, \gamma, t) = \frac{D(\alpha, t)}{D(0, t)}$, focusing for instance on the case of random transpositions $(m, q) = (2,2)$. Invariance in the class of probabilities allows for an efficient estimate of the probability distance $D$ for values of $n \lesssim 25$. Without loss of generality, we fix $\gamma = 1$ throughout the rest of this section. 

\begin{figure}
    \centering
    \subfloat[]{\includegraphics[width=0.45\linewidth]{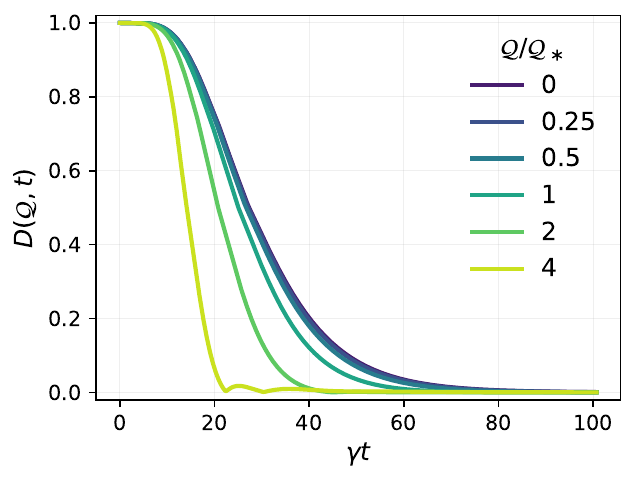}}
    \hfill
    \subfloat[]{\includegraphics[width=0.45\linewidth]{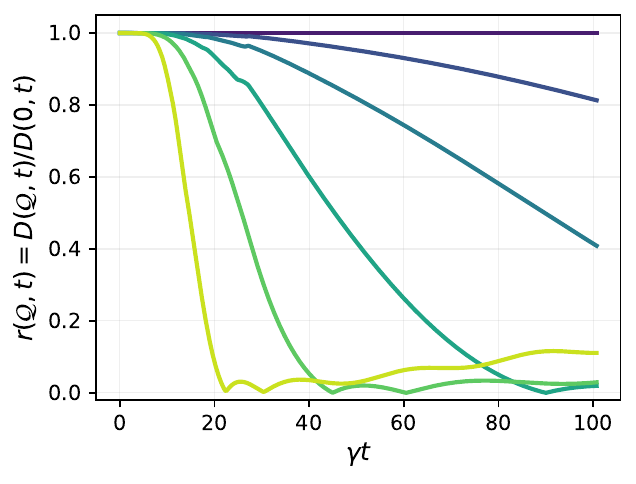}}
    \caption{Distance in probability between the QSW dynamics Eq.~\eqref{eq_proba_quantum_stochastic_walk} and the uniform distribution for random transpositions ($m = q = 2$) and $n = 23$. In (a) the distance in probability is plotted for several values of $\alpha = \mathcal{Q} \gamma$, with $\gamma = 1$. According to the criterion Eq.~\eqref{eq_Q_star}, we start to observe appreciable deviations starting from $\mathcal{Q}_* \approx 0.943$. (b) Plot of the quantum-to-classical ratio $r(\alpha, t) = D(\alpha, t)/D(0,t)$. For any value of $\alpha$ because of Eq.~\eqref{eq_bound_alpha}, the ratio is smaller than one. We observe, as $\alpha$ increases, that the first time at which the ratio deviates from unity decreases.}
    \label{f_distance_alpha}
\end{figure}

We begin by considering the $\alpha$-dependence on the probability variation distance for $n = 23$. Instead of summing over the $n!$ group elements, we utilize the fact that the probability is constant over elements of the same conjugacy class and therefore the sum is reduced to $\mathcal{P}(23) = 1255$ terms. In Fig.~\ref{f_distance_alpha}(a) the probability distance is shown for various values of $\mathcal{Q}$. The corresponding critical value is $\mathcal{Q}_* \approx 0.943$ for the threshold $\epsilon = 0.05$. Similarly Fig.~\ref{f_distance_alpha}(b) shows the quantum-to-classical ratio $r$ as a function of $\gamma t$. We observe that as $\alpha$ increases, the ratio rapidly decreases from 1 at shorter times. 

This numerical result corroborates the analytical bound Eq.~\eqref{eq_bound_alpha} and can be checked to occur with the same qualitative behavior for different choices of (even) $(m, q)$. Remarkably, the effect is actually stronger than simply providing oscillations on top of a classical exponential profile. 

We now investigate the mixing time at the threshold $\epsilon$ as a function of $\alpha$. We focus on the system sizes $n = (15, 18, 20, 23)$ and the cases $(m,q) = (2,2)$, $(2,4)$, $(4,2)$, $(4,4)$. On a common grid of $401$ values of $\mathcal{Q}=\alpha/\gamma$ and fixed $\epsilon = 0.2$, we compute the mixing time and study the quantum-to-classical ratio of mixing times $t_{\mathrm{mix}}^{(n)}(\mathcal{Q}, \epsilon)/t_{\mathrm{mix}}^{(n)}(0, \epsilon)$. Notably, this ratio is largely insensitive to the choice of $\epsilon$. When $\epsilon$ is chosen to be small, oscillations play a role and one should compute the mixing time through the first-hitting time. Our choice $\epsilon = 0.2$ is sufficiently large that oscillations do not affect the estimate of the mixing time, and that the numerical scaling of such times does not vary when $\epsilon$ is changed.

We observe that this ratio, for the $(m,q)$ considered, collapses to a common scaling function for a particular choice of variable
\begin{equation}
\label{eq:scaling}
    X_n = \frac{\mathcal{Q}}{\sqrt{a_n}}, \quad a_n = \left[\frac{2m(n-1)}{q^2 N_m} \log \left(\frac{(n-1) D_n}{n! \gamma \epsilon}\right)\right]^{-1}\underset{n \gg 1}{\sim} \frac{1}{2} \log \frac{n}{e \epsilon} \mathcal{Q}_*^2, 
\end{equation}
where we remind that $D_n$ is the number of derangements introduced in Sec.~\ref{s_classical}. We utilize a one-parameter interpolation with the function
\begin{equation}
    \frac{ t_{\rm mix}^{(n)}(\mathcal{Q},\epsilon)}{t_{\rm mix}^{(n)}(0,\epsilon)}
 = \left(1+X_n^2\right)^{-\beta}.
\end{equation}

\begin{figure}
  \centering
    \includegraphics[width=0.9\linewidth]{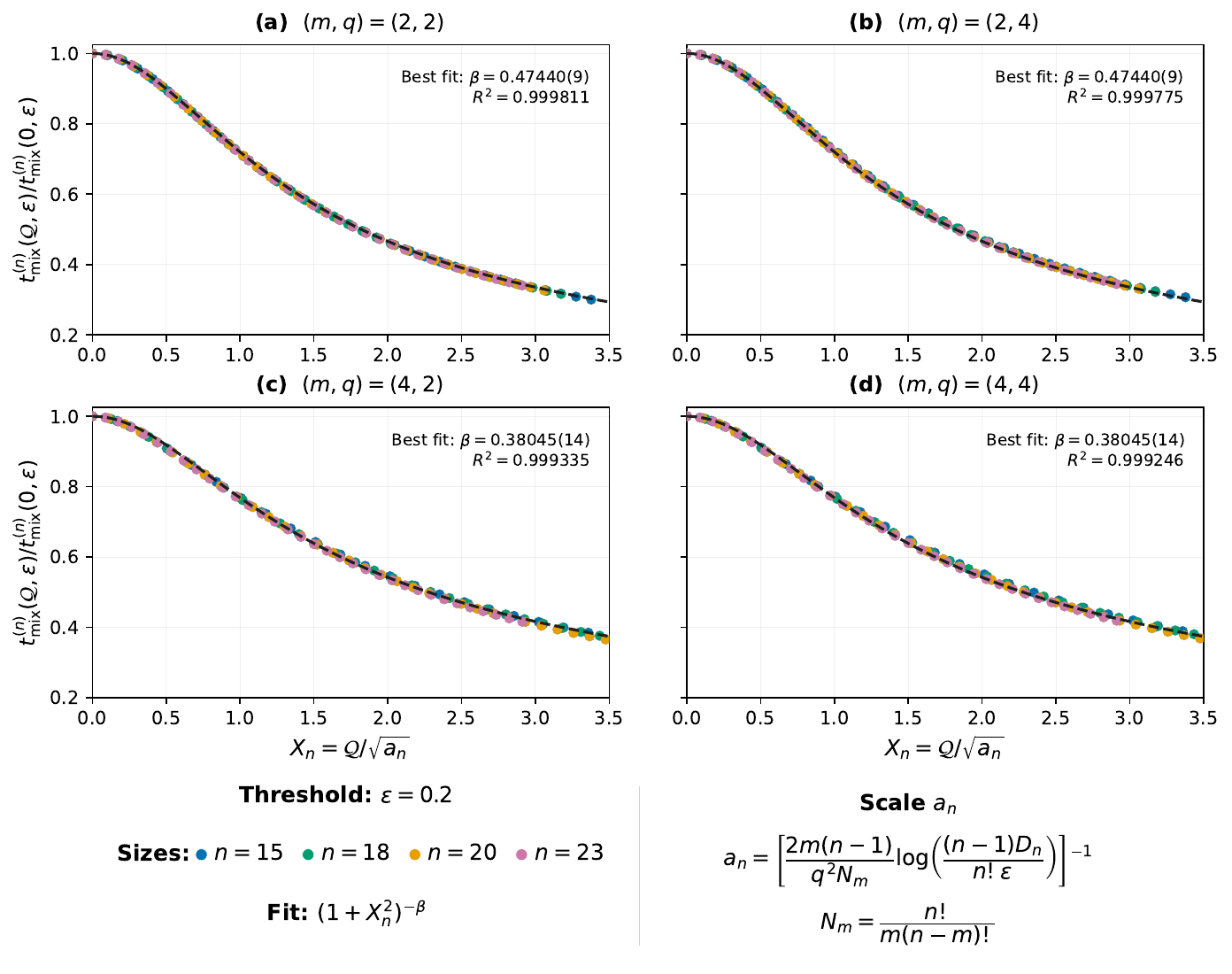}
  \caption{Quantum-to-classical ratio of mixing times for various $(m,q)$'s ((a) to (d)). The horizontal coordinate is the scaling variable $X_n = \mathcal{Q}/\sqrt{a_n}$, defined in the bottom legend and Eq.~\eqref{eq:scaling}. The sizes considered $n = \{15, 18, 20, 23 \}$ collapse to a single curve. In all cases considered, a one-parameter interpolation with the form $(1 + X_n^2)^{-\beta}$ is sufficient. The coefficient of determination of the fit is also reported.} 
  \label{f_qsw-hitting-volumes}
\end{figure}

In Fig.~\ref{f_qsw-hitting-volumes}, we report the scaling collapse and best-fit values for different $(m,q)$'s. We observe that the value of the fit parameter $\beta$ is insensitive to the particular value of $q$, but rather depends only on $m$. For $m = 2$, this value is $\beta = 0.47440(9)$ and for $m = 4$ is $\beta = 0.38045(14)$. 

The apparent independence of the exponent \(\beta\) from the dissipative cycle length \(q\) admits a natural interpretation in terms of the slow-mode decomposition. The parameter \(q\) controls the classical relaxation rate and therefore enters predominantly through the scale \(X_n\), whereas the coherent suppression factor is the characteristic function of the energy differences \(E_m(\rho)-E_m(\sigma)\) generated by the \(m\)-cycle Hamiltonian. Once the \(q\)-dependent mixing scale is absorbed into \(X_n\), the remaining shape of the crossover is therefore expected to be governed primarily by the spectral statistics of \(H_m\), and hence by \(m\). This provides a qualitative explanation for why the fitted exponent \(\beta\) is approximately independent of \(q\) for fixed \(m\). The present numerical data, however, cover only \(m,q=2,4\), and a broader analysis would be required to establish whether this behavior is universal.

We finally notice that the quadratic behavior at small \(X_n\) is consistent with the expansion of the characteristic function in Eq.~\eqref{eq_bound_kappa_s}, while the large-\(X_n\) behavior depends on higher moments of the energy-difference distribution and is not captured by the analysis of the slowest mode alone, rather it is the due to the combined effect of all the irreps entering the convolution Eq.~\eqref{eq_convolution_kernel}.

\section{Conclusions}\label{s_conclusions}

While the card shuffling problem emerges naturally in the setting of combinatorics, it admits a formulation in terms of a Markov process on the permutation group $S_n$. In this work, we focused on an essential probe of the out-of-equilibrium physics of the problem, the mixing time, originally proved by Diaconis and Shahshahani \cite{Diaconis_1981, Diaconis_1988} to be $O(n \log n)$ in the classical case. The same time-scale emerges when the continuous-time generalization of the Markov process, with rate $\gamma$, is considered.

By embedding the continuous-time random walk into a quantum stochastic dynamics, we have isolated the role played by coherence in an otherwise genuinely mixing process. The central result is the bound $D(\alpha,t)\leq D(0,t)$: although coherent evolution leaves the trace distance of the full density matrix from the maximally mixed state unchanged, it can only bring the probability distribution observed in the computational basis closer to uniformity. Coherence, therefore, does not compete with classical randomization in this setting, rather it can assist it.

The slowest irreducible mode identifies the scale of coherent coupling at which this effect becomes appreciable, while numerical results for $n \leq 23$ corroborate the analytical picture and reveal a scaling collapse of the quantum-to-classical mixing-time ratio. These results provide a particularly transparent example of how coherence and dissipation can cooperate: irreversibility supplies genuine mixing, while quantum interference accelerates the loss of classical information.

It would be interesting to investigate to what extent this mechanism can be realized in physical open quantum systems. The dissipative part of our dynamics has a natural interpretation as stochastic permutation noise, while the coherent part is generated by permutation Hamiltonians, suggesting possible connections with quantum systems possessing controllable exchange interactions. In this respect, all-to-all interacting spin systems, cold-atom platforms, and programmable quantum simulators may provide natural settings in which related dynamics could be explored. More broadly, the mechanism identified here is reminiscent of noise- or environment-assisted quantum dynamics, where coherent interference and dissipation cooperate rather than simply compete. Establishing the precise relation connecting these phenomena, and identifying experimentally realistic implementations of the permutation dynamics considered here, would be an interesting direction for future work.

\section*{Acknowledgments}
We would like to thank Ilya Chevyrev for insightful discussions. AS is grateful to Grayson Frazier for many conversations. AS's and GM's research was supported in part by grant NSF PHY-2309135 to the Kavli Institute for Theoretical Physics (KITP). 

\section*{List of Abbreviations}
In order of appearance
\begin{itemize}
    \item DS: Diaconis-Shahshahani;
    \item CCTRW: classical continuous-time random walk;
    \item QW: quantum walk;
    \item MMS: maximally mixed state;
    \item TV: total variation;
    \item irrep(s): irreducible representation(s);
    \item IPR: inverse participation ratio;
    \item RA: running average;
    \item LTA: long-time average;
    \item PH(s): permutation Hamiltonian(s);
    \item OQS: open qunatum system;
    \item QSW: quantum stochastic walk;
\end{itemize}

%%%%%%%%%%%%%%%%%%%%%%%%%%%%%%%%%%%%%%%%%%%%
%% REFERENCES
%%%%%%%%%%%%%%%%%%%%%%%%%%%%%%%%%%%%%%%%%%%%

\appendix

\section{Lower Bound on the Mixing Time of Random $q$-Cycles}\label{a_lower_bound_coupon_collector}

The following argument is the standard fixed-point lower bound for
random transpositions, adapted here to uniformly distributed
$q$-cycles. Closely related formulations appear in
\cite{Diaconis_1988}; see also
\cite{berestycki2011mixing} for the random $q$-cycle walk.

The coupon collector idea (see Section~\ref{s_DS}) can be turned into a rigorous argument that shows that, for $n\to\infty$ and for any $c > 0$
\begin{equation}
    ||p_q^{*k_-} - U|| \geq 1 - 6 e^{-c} + o(1), %\quad k_- = \frac{n}{q}(\log n - c),
\end{equation}
where $c>0$ and
\begin{equation}
\label{eq_k_lower}
    k_- = \frac{n}{q}(\log n - c).
\end{equation}
Thus, before this time, the random walk is unlikely to be mixed.

We begin by stating an equivalent definition of the DS measure
\begin{equation}
    ||p - U|| = \underset{A \subset S_n}{\max} |p(A) - U(A)|,
\end{equation}
where $A$ is a subset of $S_n$. In particular, we aim at finding an event $A$ such that $p^{* k_-}(A) \simeq 1$ and $U(A) \simeq 0$. Finding such an event would yield a \textit{lower bound} on the DS measure and therefore prove the unlikelihood of mixing at $k_-$. 

We also recall the coupon collector problem, i.e. we ask if a label $j$ has been found after $k$ applications of random $q$-cycles. Define the following indicator function for a permutation $\sigma_k = g_k \ldots g_1$
\begin{equation}
    I_j(k) = \begin{cases}
        1, \quad \text{if label $j$ has never appeared in any of the $k$ cycles $\{ g_\ell\}$}, \\
        0, \quad \text{otherwise}.
    \end{cases}
\end{equation}
Denote also by $X_k = \sum_{j = 1}^n I_j(k)$ the number of untouched labels. 

By itself $X_k$ is not sufficient to define an event. Indeed, let $\operatorname{Fix}(\sigma_k)$ the number of fixed points. i.e. the number of labels untouched by the permutation $\sigma_k$. If a label $j$ has not been touched in any of the permutations, it will be a fixed point of $\sigma_k$, therefore
\begin{equation}
\label{eq_bound_fix}
    X_k \leq \operatorname{Fix}(\sigma_k).
\end{equation}

We now prove several important properties of $X_k$:
\begin{itemize}
    \item \textbf{Average.} Since $I_j(k)$ is a boolean variable, $\mathbb{P}[I_j(k) = 1] = \mathbb{E}[I_j(k)]$ and it evaluates to 
    \begin{equation}
        \mathbb{E}[I_j(k)] = \left(1-\frac{q}{n}\right)^k = a_n^k.
    \end{equation}
    Thus, it follows that
    \begin{equation}
        \mathbb{E}[X_k] = m_n = n a_n^k.
    \end{equation}
    \item \textbf{Variance.} $I_j(k)$ is a Bernoulli variable with probability $a_n^k$. It follows that 
    \begin{equation}
        \operatorname{Var}[I_j(k)] = a_n^k (1-a_n^k) \leq a_n^k = \mathbb{E}[I_j(k)]. 
    \end{equation}
    Not only: any two labels $j_1$ and $j_2$ are not independent, rather they are correlated. The probability that two variables are not touched is
    \begin{equation}
    \begin{aligned}
        \mathbb{P}(\text{$j_1$ and $j_2$ not touched}) &= \mathbb{E}[I_{j_1}(k) I_{j_2}(k)] \\
        &= \left(\frac{\binom{n-2}{q}}{\binom{n}{q}}\right)^k = \left(\frac{(n-q)(n-q-1)}{n(n-1)}\right)^k = b_n^k.
    \end{aligned}
    \end{equation}
    Since $a_n^2 - b_n = \frac{q(n-q)}{n^2(n-1)} >0$ it follows that
    \begin{equation}
        \operatorname{Cov}[I_{j_1}(k) I_{j_2}(k)] = b_n^k - a_n^{2k} < 0. 
    \end{equation}
    Thus, such indicator functions are negatively correlated. For the total sum $X_k$, one finds
    \begin{equation}
        \begin{aligned}
            \operatorname{Var}[X_k] &= \sum_j \operatorname{Var}[I_j(k)] + 2 \sum_{j_1 < j_2} \operatorname{Cov}[I_{j_1}(k) I_{j_2}(k)]\\
            &\leq \sum_j \operatorname{Var}[I_j(k)] \leq n a_n^k = \mathbb{E}[X_k] = m_n.
        \end{aligned}
    \end{equation}
    \item \textbf{Markov inequality and Chebyschev's inequality.}
    Since the random variable $X_k$ is non-negative, one may choose an arbitrary value $0 < a \leq 1$. Such value $a$ satisfies Markov's inequality \cite{allofstatistics}
    \begin{equation}
        \mathbb{E}[X_k] \geq a \mathbb{P}(X_k \geq a).
    \end{equation}
    Similarly, we obtain Chebyschev's inequality as a special case of Markov's inequality applied to the event $|X_k - \mathbb{E}[X_k]| > a$:
    \begin{equation}
        \mathbb{P}(|X_k - \mathbb{E}[X_k]| > a) = \mathbb{P}(|X_k - \mathbb{E}[X_k]|^2 > a^2) \leq \frac{\operatorname{Var}[X_k]}{a^2}.
    \end{equation}
\end{itemize}

Using the information of the previous points, we now fix $k_- = \frac{n}{q}(\log n - c)$ and show that 
\begin{equation}
    m_n(k_-) = e^{c} + o(1).
\end{equation}
Application of Chebyschev's inequality to the value $a = m_n(k_-)/2$ gives
\begin{equation}
\mathbb{P}(X_k < m_n/2)  \leq 
    \mathbb{P}(|X_k - m_n| > m_n/2)  \leq 4\frac{\operatorname{Var}[X_k]}{m_n^2} \leq \frac{4}{m_n} = 4 e^{-c}.
\end{equation}
Conversely, it is very likely that the variable $X_k$ is larger than its half-mean when $c> 0$.

We are now ready to construct the event, which is
\begin{equation}
    A_{n,c} = \set{\sigma \in S_n| \operatorname{Fix}(\sigma) \geq \frac{m_n}{2}}.
\end{equation}
From Eq.~\eqref{eq_bound_fix} it follows that 
\begin{equation}
    \set{X_k \geq \frac{m_n}{2} } \subseteq A_{n,c},
\end{equation}
and therefore 
\begin{equation}
    p_q^{*k_-}(A_{n,c}) \geq \mathbb{P}\left({X_k \geq \frac{m_n}{2} }\right) \geq 1 - 4 e^{-c} + o(1).
\end{equation}
This tells that the event is likely to happen. 

What about the same event under the uniform distribution? The probability that a uniformly chosen permutation does not involve site $j$ is $1/n$ -- there are $(n-1)!$ permutations which leave $j$ untouched out of $n!$. Therefore the average number of fixed points for a permutation drawn from the uniform distribution is
\begin{equation}
    \mathbb{E}_U[\operatorname{Fix}] = n \frac{1}{n} = 1.
\end{equation}
Applying Markov's inequality we have
\begin{equation}
    U_{q, k_-}(A_{n,c}) = \mathbb{P}\left(\operatorname{Fix}\geq \frac{m_n}{2}\right) \leq \frac{2}{m_n} = 2 e^{-c} + o(1).
\end{equation}

These two considerations can be summarized in the following lower bound for the DS distance
\begin{equation}
    ||p_q^{k_-} - U || \geq p_q^{k_-}(A_{n,c})- U_{q, k_-}(A_{n,c}) \geq 1 - 6 e^{-c} + o(1).
\end{equation}
Thus, it is unlikely that for any finite $c>0$, when $n\to\infty$, the random walk can mix. 

Thus, for any fixed sufficiently large $c>0$, the walk is still far from equilibrium at 
\begin{equation*}
k = \frac{n}{q}(\log n - c).
\end{equation*}
Consequently, to leading order in $n$,
\[
t_{\rm mix}(q)\gtrsim \frac nq\log n . \]

\section{IPR of Class Operators}\label{a_IPR_class}

In Section~\ref{s_classical} and Section~\ref{s_quantum} we argued that class operators can be exactly solved in terms of the characters of $S_n$. In this Appendix, we shall show that a statistical analysis of such characters can inform about the value of the quantum dephasing time for the unitary QW.

In the main part of the manuscript, we defined QWs through Hamiltonians which coincide with the transition matrix of a CCTRW. We first observe how the dynamical properties of the two dynamics are related to different spectral regions of the operator $H$. 

For the CCTRW, long-time relaxation is governed by the spectral outliers closest to the stationary eigenvalue. For the unitary quantum walk, decoherence is driven by interference among a macroscopic number of energy levels; thus, it is more susceptible to the statistics of the bulk spectrum of the Hamiltonian. This distinction becomes particularly transparent when computing the IPR.

Indeed, we recall that for the class operators $H_q$, the IPR of the QW reads
\begin{equation}
\label{eq_IPR_class_quantum}
    \mathrm{IPR}(t) = \sum_{g \in S_n} \left|\sum_{\mu} \frac{d_\mu}{n!} \chi_\mu(g) e^{- i E_q(\mu) t} \right|^4, \qquad E_q(\mu) = \frac{\chi_\mu^{[q]}}{d_\mu},
\end{equation}
where the sum runs over all the irreps of $S_n$. The sum can be expanded and reads
\begin{equation}
\begin{aligned}
    \mathrm{IPR}(t) &= \sum_{\mu_1, \ldots, \mu_4} \frac{d_{\mu_1}d_{\mu_2}d_{\mu_3}d_{\mu_4}}{(n!)^4}\left( \sum_{g \in S_n} \chi_{\mu_1}(g) \chi_{\mu_2}(g) \chi_{\mu_3}(g) \chi_{\mu_4}(g)\right) \\
    &\times \exp\left(-i t \left(\frac{\chi_{\mu_1}^{[q]}}{d_{\mu_1}} - \frac{\chi_{\mu_2}^{[q]}}{d_{\mu_2}}+\frac{\chi_{\mu_3}^{[q]}}{d_{\mu_3}}- \frac{\chi_{\mu_4}^{[q]}}{d_{\mu_4}}\right)\right).
\end{aligned}
\end{equation}
The dependence on the group element $g$ can be eliminated by introducing the conventional inner product on the group \cite{elliottdawber}:
\begin{equation}
    \braket{f, h}_{S_n} = \frac{1}{n!} \sum_{g \in S_n} f(g) h^*(g).
\end{equation}
Then $\sum_{g \in S_n} \chi_{\mu_1}(g) \chi_{\mu_2}(g) \chi_{\mu_3}(g) \chi_{\mu_4}(g) = n! \braket{\chi_{\mu_1}\chi_{\mu_2}, \chi_{\mu_3}\chi_{\mu_4}}_{S_n}$. However, the individual functions in the inner product admit a representation in terms of the Kronecker coefficients of $S_n$:
\begin{equation}
    \chi_{\mu}(g) \chi_{\nu}(g) = \sum_\rho g_{\mu \nu}^{\rho} \chi_{\rho}(g),
\end{equation}
and therefore the sum can be written compactly as
\begin{equation}
    \frac{1}{n!}\sum_{g \in S_n} \chi_{\mu_1}(g) \chi_{\mu_2}(g) \chi_{\mu_3}(g) \chi_{\mu_4}(g) = \sum_\rho g_{\mu_1 \mu_2}^\rho g_{\mu_3 \mu_4}^\rho,
\end{equation}
after applying the character orthogonality relations \cite{elliottdawber}.

Define $\Omega$ as the set of distinct energy differences. For each value $\Delta$, we have
\begin{equation}
    \Omega_{\Delta} = \set{\mu_1, \mu_2, \mu_3, \mu_4 | E_{\mu_1} - E_{\mu_2} + E_{\mu_3} - E_{\mu_4} = \Delta}. 
\end{equation}
The IPR can then be rewritten as
\begin{equation}
\begin{aligned}
    \mathrm{IPR}(t) &= \frac{1}{n!} \sum_{\Delta \in \Omega} c_\Delta \cos (\Delta t),\\
    c_\Delta &= \sum_{\mu_1, \ldots, \mu_4} \frac{d_{\mu_1} d_{\mu_2} d_{\mu_3} d_{\mu_4}}{(n!)^2} \sum_\rho g_{\mu_1 \mu_2}^\rho g_{\mu_3 \mu_4}^\rho, \\
    \Delta &= \frac{\chi_{\mu_1}^{[q]}}{d_{\mu_1}} - \frac{\chi_{\mu_2}^{[q]}}{d_{\mu_2}}+\frac{\chi_{\mu_3}^{[q]}}{d_{\mu_3}}- \frac{\chi_{\mu_4}^{[q]}}{d_{\mu_4}}.
\end{aligned} 
\end{equation}

In terms of $\Delta$, large $\Delta$'s correspond to fast oscillatory modes, while smaller $\Delta$'s are responsible to long-time dephasing. To quantify such separation of time-scales, one can study, instead of the instantaneous IPR, its smearing 
\begin{equation}
    \braket{\operatorname{IPR}(t)}_\tau = \frac{1}{\tau} \int_{t-\tau/2}^{t + \tau/2} \dd T'\; \operatorname{IPR}_2(T') = \frac{1}{n!} \sum_{\Delta \in \Omega} c_{\Delta}\, \operatorname{sinc}\left(\frac{\Delta \tau}{2}\right)\, \cos({\Delta t}),
\end{equation}
where $\operatorname{sinc}(x) = \sin(x)/x$. Setting $\tau = 1$ and $T = k \in \{0,1,2,\ldots\}$ one reduces the continuous-time dynamics of the QW to a discrete-time one. In practice, one applies random measurements on an interval centered at integer $t$ with unit width. The $\operatorname{sinc}$ function offers smooth regularization, suppressing oscillations at frequencies $\Delta = O(1)$ while preserving slower modes.

For $H_q$, the $\operatorname{IPR}$ is the sum of two contributions: the LTA $\overline{\operatorname{IPR}} = c_0/ n!$, which consists of the irreps satisfying $E_{\mu_1} + E_{\mu_3} = E_{\mu_2} + E_{\mu_4}$ and the off-diagonal $\widetilde{\operatorname{IPR}}(t)$ one. The dephasing time $t_{\text{deph}}$ can be computed through the first zero of $\widetilde{\operatorname{IPR}}(t)$, i.e. the first-hitting time of the LTA. We now show how the spectral analysis of the modes allows to estimate the dephasing time Eq.~\eqref{eq_QW_dephasing} of the main text.

The treatment of the $\operatorname{IPR}$ simplifies in the thermodynamic limit, $n\to\infty$. There, irreps $\rho$ can be treated as random variables distributed according to the Plancherel measure $\dd \mu[\rho] = d_\rho^2 /n!$ and $t_U$ can be extracted with statistical arguments as follows. The $\Delta$ entering $\operatorname{IPR}$ can be divided in modes with a definite finite-$n$ scaling:
\begin{itemize}
    \item \textbf{Standing modes}, $\Delta = O(1)$. A particular representative is $\Delta = 0$, e.g. when $\rho = \rho'$ and $\sigma = \sigma'$. These contribute to the LTA, which is time-independent.
    
    \item \textbf{Bulk modes}. These modes are typical and characterize the \textit{bulk} of the Plancherel measure: diagrams whose dimension is $d_\rho \sim \sqrt{n!/\mathcal{P}(n)}$. The particular shape of the distribution was derived by Kerov \cite{kerov_1993}. The $q$-class characters follow a central-limit theorem \cite{Kerov_central_limit_theorem_characters}: $\frac{\chi_\rho^{[q]}}{d_\rho} \overset{L\to\infty}{\longrightarrow} \mathcal{N}\left(0, q n^{-q}\right)$. Hence, typical gaps are of order $O(1/\sqrt{n}^q)$ for $H_q$. 
    
    \item \textbf{Outlier modes}. For $q$-class PH, these occur when the representations are close to the trivial one and populate the tails of the Plancherel measure. For the particular case of $\rho = [n]$ and $\rho' = [n-1, 1]$, their gap is $\delta_{\rho \rho'} = q/(n-1) = O(1/n)$. However, outlier modes become superexponentially suppressed in the Plancherel measure. Indeed, from the hook formula, their dimension is polynomial rather than factorial (bulk). 
\end{itemize}

The dephasing time obtained from the statistical behavior of eigenvalues in the bulk gives the quantum dephasing time for class operators, as defined in Section~\ref{s_quantum}. However, it neglects the structure of the eigenvalue coefficients $c_\Delta$. A statistical approach to the evaluation of the IPR, accounting for the correlation between eigenvectors and eigenvalues, is an interesting research direction of interest in the study of nonequilibrium properties in integrable and chaotic quantum many-body systems.

\begin{figure}
    \centering
    \includegraphics[width=0.45\linewidth]{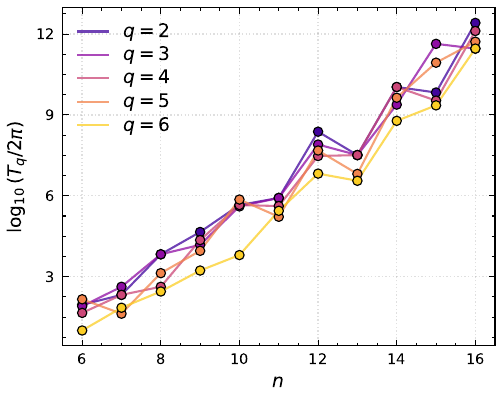}
    \caption{Dependence of the period $T_{q}(n)$ of $q$-class PHs with the number of sites $n$. This period likely scales as $T_{q}(n) = O(e^n)$.}
    \label{f_period}
\end{figure}

\paragraph{Recurrence Time.} To conclude this Appendix, we briefly discuss an important property of such class operators, i.e. the period of the unitary generator $e^{- i H_q t}$. The energy associated to each irrep $\rho$ is \textit{rational}: $E_\rho^{[q]} = \chi_{\rho}^{[q]}/d_\rho$. This implies that the unitary time evolution is periodic of time $T_{[q]}(n)$. The period is a function of the characters, and can be found as follows:
\begin{itemize}
    \item first, the irreps with nonzero character are selected. Let this set be denoted as $\{\rho\}^* = \{ \rho \text{ s. t. } \chi_{\rho}^* \neq 0 \}$. Indeed, if a character is zero, so is its associated energy. Zero energies do not generate interference patterns.
    \item Second, we compute the least common multiple (LCM) of the $d_{\rho}$'s and the greatest common divisor (GCD) of the $\chi_{\rho}^{[k]}$ in $\{\rho\}^*$.
    \item The period is finally given by
    \begin{equation}
        T_{[q]}(n) = 2 \pi \frac{\text{LCM}(d_\rho^*)}{\text{GCD}(\chi_\rho^{[q],*})}.
    \end{equation}
\end{itemize}
In Fig.~\ref{f_period} we compute such period for $n\in [6, 16]$, which is likely of order $ T_{[q]}(n) = O(e^n)$. This is confirmed by seemingly straight lines in the log-plot. This result is consistent with the law $\sum_\rho d_\rho^2 = n!$ in the regular representation, meaning that typical bulk irreps have dimension $d_\rho \sim n!/\mathcal{P}(n)$, where $\mathcal{P}(n)$ is the number of partitions of $n$, coincident with the number of irreducible representations. We also observe that for moderate systems sizes $n \geq 10$, the order of magnitude of such recurrence time is $10^5$. 

\printbibliography

\end{document}